\documentclass[a4paper,11pt]{article}
\pdfoutput=1
\usepackage{jheppub}
\usepackage[utf8x]{inputenc}
\usepackage{bm}
\usepackage{float}
\usepackage{tikz}
\usepackage[nottoc,notlof,notlot]{tocbibind} 

\graphicspath{{figures/}}
\hypersetup{unicode=true}

\PrerenderUnicode{⁺}\PrerenderUnicode{⁻}
\PrerenderUnicode{λ}\PrerenderUnicode{¹}

\newcommand{\be}{\begin{equation}}
\newcommand{\ee}{\end{equation}}

\newcommand\fig[2] {\begin{figure}[#1]\centering #2\end{figure}}

\newcommand{\wh}{\widehat}

\newcommand{\one}{{\rm 1\kern -.9mm l}}

\newcommand{\cpone}{\mathbb{CP}^1}

\newcommand{\bB}{{\mathbb B}}

\numberwithin{equation}{section} 
\begin{document}

\title{\boldmath Exact WKB analysis of $\cpone$ holomorphic blocks}

\author[a]{Sujay K. Ashok,}
\affiliation[a]{Institute of Mathematical Sciences,\\
Homi Bhabha National Institute (HBNI),\\
IV Cross Road, C.~I.~T.~Campus, Taramani,\\
Chennai 600113, India\\}
\emailAdd{sashok@imsc.res.in} 
 
\author[a]{P. N. Bala Subramanian,}
\emailAdd{pnbala@imsc.res.in}

\author[b]{Aditya Bawane,} 
\affiliation[b]{Department of Physics,\\
Indian Institute of Technology Madras,\\
Chennai 600036, India\\}
\emailAdd{ph18ipf07@smail.iitm.ac.in}

\author[c]{Dharmesh Jain,}
\affiliation[c]{Theory Division, Saha Institute of Nuclear Physics,\\
1/AF Bidhan Nagar, Kolkata 700064, India\\}
\emailAdd{d.jain@saha.ac.in}

\author[d]{\mbox{Dileep P. Jatkar,}} 
\affiliation[d]{Harish-Chandra Research Institute,\\
Homi Bhabha National Institute (HBNI),\\
Chhatnag Road, Jhusi,\\ 
Allahabad 211019, India\\}
\emailAdd{dileep@hri.res.in}

\author[a]{Arkajyoti Manna}
\emailAdd{arkajyotim@imsc.res.in}

\abstract{We study holomorphic blocks in the three dimensional ${\mathcal N}=2$ gauge theory that describes the $\cpone$ model. We apply exact WKB methods to analyze the line operator identities associated to the holomorphic blocks and derive the analytic continuation formulae of the blocks as the twisted mass and FI parameter are varied. The main technical result we utilize is the connection formula for the ${}_1\phi_1$ $q$-hypergeometric function. We show in detail how the $q$-Borel resummation methods reproduce the results obtained previously by using block-integral methods.
}


\maketitle

\section{Introduction and summary}

Three-dimensional supersymmetric gauge theories are known to
exhibit interesting dynamics, such as mirror symmetry, and IR dualities 
\cite{Intriligator:1996ex,Aharony:1997bx,Dorey:1999rb,
Tong:2000ky}. With the application of
localization methods \cite{Witten:1988ze,Pestun:2007rz} to these theories on
(squashed) $S^3$ \cite{Kapustin:2009kz,Jafferis:2010un,Hama:2010av,Hama:2011ea,Imamura:2011wg,Nian:2013qwa}, it
became possible to compute their exact partition functions and other
supersymmetric observables. This opened up new avenues to further delve into
the rich dynamics of these theories and uncover possibly new 
symmetries and dualities, including holography (see chapters 6-8 of
\cite{Pestun:2016jze} and references therein). A new perspective on
computing these partition functions was discovered shortly afterwards
\cite{Pasquetti:2011fj,Dimofte:2011py} in terms of holomorphic
blocks. They were then extensively studied in \cite{Beem:2012mb} as
fundamental objects using which the partition functions and (twisted)
indices of these 3d gauge theories can be obtained by gluing these
blocks in distinct ways. These have proved useful in understanding
mirror symmetry and discovering more dualities (see, for example,
\cite{Zenkevich:2017ylb,Aprile:2018oau}). The factorization into
holomorphic blocks have been shown for partition functions on more
general 3-manifolds \cite{Imamura:2013qxa,Nieri:2015yia,Closset:2017zgf,
Closset:2018ghr,Pittelli:2018rpl}. They have also played a key role in the 3d/3d
correspondence in which they were mapped to partition functions of
complex Chern-Simons theories on Lefschetz
thimbles \cite{Witten:2010cx,Dimofte:2010tz,Witten:2010zr,
Witten:2011zz,Dimofte:2011ju}. The 3d holomorphic blocks are also
related to 4d and 5d theories where such a factorization of partition
functions is again
observed \cite{Yoshida:2014qwa,Nieri:2015yia,Pasquetti:2016dyl,Longhi:2019hdh}.

One of the main properties of the 3d holomorphic blocks that we focus
on in this paper is that they are solutions to $q$-difference
equations.  These are referred to as line operator identities (LOIs)
in \cite{Beem:2012mb} and can be derived systematically given the
ultraviolet description of the gauge theory.  Another important
property of interest to physical applications is that these blocks
exhibit Stokes phenomena as the parameters of the gauge theory are
varied.  The physical parameters are the complexified masses and FI
parameters of the gauge theory.  The parameter space is divided up
into Stokes regions and in each such region, the LOIs have as many
independent solutions as the number of massive vacua of the gauge
theory.  Since the blocks solve linear difference equations, in each
Stokes region the holomorphic blocks form a basis; this basis can be
written as a linear combination of the basis in the neighbouring Stokes regions.
The Stokes matrices/multipliers give the relation between these bases
defined in the Stokes regions separated by a Stokes line.
In \cite{Beem:2012mb}, it was also shown that the holomorphic blocks
could be written as finite dimensional contour integrals (termed
block-integrals) such that they automatically solve the LOIs.
The Stokes phenomenon exhibited by the holomorphic blocks
was then shown to be a consequence of a change of contours as the Stokes lines are crossed.

In this work, we approach the same problem from a purely algebraic
perspective and show that the Stokes behaviour of the blocks can be
obtained by analyzing the exact WKB properties of the $q$-difference
equations that are satisfied by the holomorphic blocks (see \cite{RSZ} for an introduction to $q$-difference equations).
We focus on the $\cpone$ model \cite{Witten:1993yc,Dorey:1999rb,Tong:2000ky},
which has a gauge theory description in the ultraviolet as a U$(1)$
gauge theory with two charged chiral multiplets.  
This is the simplest model in which the LOIs have an irregular singular
point, in addition to a regular singular point \cite{TAB}. 
Up to prefactors that are given in terms of
$\Theta_q$-functions, the blocks near the regular singular point are
given in terms of the $q$-hypergeometric function ${}_1\phi_1(0;a;q,z)$\footnote{For this $q$-hypergeometric function, the regular singular point is at $z=0$ and the irregular singular point is at $z=\infty$.}.

We cannot directly solve these LOIs near the
irregular singular point and have to turn to connection formulae,
which relate solutions of $q$-difference equations near different
singular points.  However, the well-known connection formulae
\cite{Watson:1910ghs} relate the solutions near two regular
singularities.  Whereas these solutions have finite radius of
convergence, solutions around an irregular singularity are typically
asymptotic series with zero radius of convergence. To extend the
connection formulae to solutions near irregular singularities, one has
to augment the procedure by first carrying out the $q$-Borel summation
\cite{Morita:2011hx,MOR,DRELO,OHY,OHYtalk,Adachi} of the asymptotic series near the
irregular singular point. These methods have been applied to the
study of the connection formula for ${}_1\phi_1$-function in $|q|<1$
chamber but we also need the connection formula in the $|q|>1$ chamber
to completely characterize the Stokes phenomena exhibited by
$\cpone$ model.  One of the main technical results of this work is a
derivation of the $|q|>1$ formula suitably adapting the treatment of the
$|q|<1$ result in \cite{OHY}. 

An important subtlety in the derivation of the connection formulae is that the analytic continuation of the ${}_1\phi_1$-function in fact depends on the choice of an arbitrary complex number $\lambda$.  Naively it would appear as if there is a one-parameter family of analytically continued holomorphic blocks.  However, there are {\it two} independent LOIs in the $\cpone$ model and it turns out that it is only for two particular choices of $\lambda$, determined by the physical parameters of the theory, that the analytic continuation leads to consistent holomorphic blocks.  So, in the end, this procedure leads to three pairs of $\cpone$ blocks (one set near the regular singular point and two sets near the irregular singular point) and each of these pairs correspond to a basis in a particular Stokes region in the parameter space.  These correctly reproduce the expected Stokes behaviour of the $\cpone$ blocks, derived from the block-integral analysis in \cite{Beem:2012mb}.

This paper is organized as follows: In Section \ref{qborel} we review
the $q$-Borel and $q$-Laplace transforms to solve the $q$-difference
equations.  In Section \ref{cponemodel} we review the $\cpone$
model, obtain the holomorphic blocks near the regular singular point by
solving the LOIs explicitly, and briefly review the results of
\cite{Beem:2012mb}.  Then in Section \ref{qstokesCP1} we apply the
results of Section \ref{qborel} to write down the connection formulae
that relate holomorphic blocks in the different Stokes regions. Finally, in Section \ref{Stokesfromconn} we bring all the results together to identify the pair of holomorphic blocks in each Stokes region along with explicitly identifying the relevant regions in the parameter space of the $\cpone$ model.
We also have three technical appendices, including the detailed derivation
of the connection formula in $|q|>1$ chamber for the ${}_1\phi_1$-function
in Appendix \ref{proofofconn}.

\section{\texorpdfstring{$\bm{q}$-Borel resummation for
    $\bm{q}$-difference equations}{q-Borel resummation for
    q-difference equations}}\label{qborel} 

Our goal in this work is to study holomorphic blocks in various
regions of parameter space and to analyze how these blocks behave as
one crosses Stokes lines in the parameter space using purely algebraic
techniques.  The holomorphic blocks obey line operator identities, which are a set of $q$-difference
equations. In this section, following \cite{MOR, OHY}, we review the $q$-Borel
resummation methods that allow one to eventually solve the connection problem
of analytically continuing solutions of $q$-difference equations around an irregular 
singular point to solutions around a regular singular point.

\subsection[\texorpdfstring{The $q$-Borel transform}{The q-Borel
  transform}]{The $\bm{q}$-Borel transform} 

We begin with a $q$-difference equation of the form
\be
{\mathcal D} [\sigma_q(t)]\, f(t) = 0\,,
\ee
where $\sigma_q(t) = q^{t\frac{d}{dt}}$ such that $\sigma_q(t)f(t)=f(qt)$ and we look for solutions near
the point $t=0$. The $q$-Borel resummation involves two steps: i)
$q$-Borel transform followed by ii) its inverse, the $q$-Laplace
transform.  The $q$-Borel transform of a formal series
$f(t)=\sum_{n=0} a_n t^n$ is defined for both $|q|\gtrless 1$ as
follows:
\begin{equation}
\mathcal{B}^\pm_q [f(t)](\tau)=\sum_{n=0}^\infty a_n q^{\pm\frac{n}{2} (n-1)}\tau^n \,.
\end{equation}
In what follows, we will apply this operator to the $q$-difference
equation and the following result will prove useful:
\begin{equation}
\label{Bplusminusids}
\mathcal{B}^\pm_q \left[t^m \sigma^p_q(t) f(t)\right](\tau)= q^{\pm \frac{m}{2}(m-1)}\tau ^m \sigma_q^{p\pm m} (\tau)\mathcal{B}^\pm_q [f(t)](\tau) \,.
\end{equation}

\subsection[\texorpdfstring{{The $q$-Laplace transform}}{The q-Laplace
  transform}]{The $\bm{q}$-Laplace transform}\label{sec:qLaplaceT}

After acting with the $q$-Borel transform on a divergent solution
around an irregular singular point, we use the inverse transform to get
the actual solution. There are two types of $q$-Borel transforms and
there are correspondingly two types of inverse transforms which we
discuss in turn, following \cite{OHY}. We will also see in the following sections that as the
theta function $\Theta_q(x)$ has different series expansion for different $|q|$
chambers, we have to use different $q$-Laplace transforms in the corresponding $q$-chambers.

\subsubsection[\texorpdfstring{The $q$-Laplace transform
  $\mathcal{L}_{q}^{-}$ for $|q|<1$}{The q-Laplace transform Lq⁻ for
  |q|<1}]{The $\bm{q}$-Laplace transform $\bm{\mathcal{L}_{q}^{-}}$
  for $\bm{|q|<1}$}

We define the $q$-Laplace transform $\mathcal{L}_q^{-}$ to be given by the
contour integral \cite{OHY}:
\begin{equation}
\label{qLapminus}
\mathcal{L}_q^{-}[f(t)](\tau) =\frac{1}{2\pi i}
  \oint_{\Gamma_{\epsilon}} \frac{ds}{s}f(s)
  \Theta_q\left(\tau s^{-1}\right),
\end{equation}
where the contour $\Gamma_{\epsilon}$ is a circle of small radius $\epsilon$ in the
complex $s$-plane around $s=0$ and the $\Theta_q$-function is defined in Appendix \ref{specialfunctions}. Consider the action of this operation on a convergent power series $g(z)$. Any such power series is written in the form:  
\be
g(z)=\sum _{n \geq 0} a_n b_n z^n\,.
\ee
If the series has a finite radius of convergence (say $r$), it can be
re-expressed in terms of the following integral (using Cauchy's
residue theorem):
\begin{equation}
\sum_{n \geq 0} a_n b_n z^n =\frac{1}{2\pi i} \oint_{\Gamma
  _{\epsilon}} \frac{ds}{s} \alpha (s) \beta\left(zs^{-1}\right),
\label{l2}
\end{equation}
where $\alpha (s)=\sum_{n\geq 0} a_n s^n$ and
$\beta (zs^{-1})=\sum_{m\geq 0} b_m (zs^{-1})^m$. If $\alpha(s)$ and $\beta(s)$ are two convergent
series with maximum radius of convergence $r$, then we see that equation \eqref{l2} holds. 

Let us now set
\be
\alpha(s) =\mathcal{B}_q^-[f(t)](s) \quad \text{ and }\quad \beta\left(zs^{-1}\right) = \Theta_q\left(-zs^{-1}\right),
\ee
and calculate the r.h.s of \eqref{l2}:
\begin{equation}
\frac{1}{2\pi i} \oint_{\Gamma_{\epsilon}} \frac{ds}{s} \sum_{n\geq 0}
  \sum_{m\in \mathbb{Z}} a_n q^{-\frac{n}{2}(n-1)}
  q^{\frac{m}{2}(m-1)}z^m s^{n-m}\,.
\end{equation}
In the last summation one can restrict the values in the summation
over $m$ to those with $m \geq 0$ as the $m\leq -1$ values do not give
rise to poles. Hence, for $m,n\geq 0$ the non-zero contributions arise only from
the case $n=m$ (as all higher order residues are vanishing), which lead to the following result:
\begin{equation}
\frac{1}{2\pi i} \oint_{\Gamma_\epsilon}
  \frac{ds}{s}\mathcal{B}_q^-[f(t)](s)
  \Theta_q\left(-zs^{-1}\right) =\sum_{n\geq 0} a_n  z^n =f(z)
  \,.
\end{equation} 
So, we find the $q$-Laplace transform defined in \eqref{qLapminus} inverts the $q$-Borel transform ${\mathcal B}^-_q$ for convergent power series:
\begin{equation}
(\mathcal{L}_q^{-} \circ \mathcal{B}^-_q) [f]=f\,. \label{l1}
\end{equation}

While applying this formalism to find the holomorphic blocks of the $\cpone$ model, we will begin with a particular ansatz for the solution of a $q$-difference equation that involves factoring out $\Theta_q$-function (or its inverse) from a formal power series (see \eqref{h1ansatz}, \eqref{h2ansatz}). Then, applying the $q$-Borel transform to this power series, we will see that ${\mathcal B}_q^-[f](s)$ has simple poles in the $s$-plane. Applying the $q$-Laplace transform ${\mathcal L}^-_q$ as defined in \eqref{qLapminus}, we shall see that deforming the contour to pick up these poles gives rise to a connection formula.

\subsubsection[\texorpdfstring{The $q$-Laplace transform
  $\mathcal{L}_{q,\lambda}^{+}$ for $|q|<1$}{The q-Laplace transform
  L\{q,λ\}⁺ for |q|<1}]{The $\bm{q}$-Laplace transform
  $\bm{\mathcal{L}_{q,\lambda}^{+}}$ for $\bm{|q|<1}$}

The $q$-Laplace transform of type $+$ is defined as follows \cite{OHY}:
\be
\mathcal{L }_{q,\lambda}^{+} [f(\tau)](t) =\sum_{m\in \mathbb{Z}} \frac{f(\lambda q^m)}{\Theta_q\left(-\lambda q^m t^{-1}\right)} \,\cdot
\label{LqPluslam}
\ee
Note that this transform depends on an extra complex parameter $\lambda$. With this definition one can show that both $q$-Borel and $q$-Laplace transforms are additive under addition of different functions. Using
this fact we can show that
\begin{equation}
(\mathcal{L }_{q,\lambda}^{+} \circ \mathcal{B}_q^{+})[f] =f\,, \label{A.4-1}
\end{equation}
where $f(x)$ is convergent.  We present the inductive proof following
\cite{OHY}.  First we note that 
\begin{equation}
\mathcal{L}_{q,\lambda}^+ [1](t)=\sum_{m\in
  \mathbb{Z}}\frac{1}{\Theta_q \left(-\lambda
  q^m t^{-1}\right)}=\frac{1}{\Theta _q
  \left(-\lambda t^{-1}\right)}\sum_{m\in \mathbb{Z}} \lambda^m
  t^{-m}q^{\frac{m}{2}(m-1)}=1\,.
\end{equation}
We always set $a_0 =1$ in formal series so that we can have
\begin{equation}
\mathcal{B}^+_q [1](\tau) =1\,.
\end{equation}
Now we assume that 
\begin{equation}
\mathcal{L}_{q,\lambda}^+\big[ \mathcal{B}^+_q [a\, z^n](\tau)\big](t)=a\, t^n \,.
\end{equation}
By noting the following relation:
\begin{equation}
\mathcal{B}^+_q \big[a\, z^{n+1}\big](\tau)=q^n \tau \mathcal{B}^+_q [a\, z^n](\tau)\,,
\end{equation}
one can show that 
\begin{equation}
\mathcal{L}_{q,\lambda}^+\big[ \mathcal{B}^+_q \big[a\, z^{n+1}\big](\tau)\big](t)=a\, t^{n+1} \,.
\end{equation}
Thus, if the function $f(t)=\sum _{n}a_n t^n$ with ($a_0 =1)$ is
convergent then we have a proof of \eqref{A.4-1}.  The key point to
note here is the choice of a complex number $\lambda$ in the
definition of the $q$-Laplace transform \cite{OHY, DRELO}. We first denote by $[\lambda]$ an equivalence class of $\lambda$ in  $\mathbb{C}^{\star}/q^\mathbb{Z}$. We now have the constraint $\left[\lambda\right] \in \left(\mathbb{C}^{\star}/q^\mathbb{Z}\right)\setminus \{\left[1\right] \}$ since if $\lambda$ is an integral power of $q$, the definition in \eqref{LqPluslam} leads to a divergent result. The choice of $\lambda$ will prove to be  important in providing different ways to analytically continue the holomorphic blocks.

\subsubsection[\texorpdfstring{The $q$-Laplace transform
  $\mathcal{L}_{q}^{+}$ for $|q|>1$}{The q-Laplace transform Lq⁺ for
  |q|>1}]{The $\bm{q}$-Laplace transform $\bm{\mathcal{L}_{q}^{+}}$
  for $\bm{|q|>1}$}

We define this $q$-Laplace transform for $|q|>1$ as follows
\begin{equation}
\mathcal{L}_q^{+}[f(t)](\tau) =\frac{1}{2\pi i}\oint_{\Gamma_{\epsilon}} \frac{ds}{s}f(s) \Theta_q^{-1}\left(-q \tau s^{-1}\right).
\label{qlaplace1}
\end{equation}
Notice that this definition follows naturally from the definition \eqref{qLapminus} for 
$|q|<1$, in view of the following transformation property of theta
function
\begin{align}
\Theta_q(x)=\Theta_{q^{-1}}^{-1}(x^{-1})\,.
\end{align}
 This definition is consistent since the theta function  has the
 following series expansion for $|q|>1$:
\begin{align}
\Theta_{q} ^{-1}\left(-q \tau s^{-1}\right) =\sum_{n \in \mathbb{Z}}
  q^{-\frac{n}{2}(n-1)} \left(\tau s^{-1}\right)^n\,.
\end{align} 
We denote this $q$-Laplace transform as $\mathcal{L}_q^{+}$ because now
it satisfies
\begin{align}
(\mathcal{L }_{q}^{+} \circ \mathcal{B}_q^{+})[f] =f\,.
\end{align}

\subsubsection[\texorpdfstring{The $q$-Laplace transform
  $\mathcal{L}_{q,\lambda}^{-}$ for $|q|>1$}{The q-Laplace transform
  L\{q,λ\}⁻ for |q|>1}]{The $\bm{q}$-Laplace transform
  $\bm{\mathcal{L}_{q,\lambda}^{-}}$ for $\bm{|q|>1}$}

The $q$-Laplace transform of type $-$ for $|q|>1$ can be defined as follows:
\be
\mathcal{L }_{q,\lambda}^{-} [f(\tau)](t) =\sum_{m\in \mathbb{Z}} f(\lambda q^m)\,\Theta_q\left(-\lambda q^{m+1} t^{-1}\right).
\ee
Its consistency can again be checked by following the previous analysis for $|q|<1$.

\section{\texorpdfstring{The $\bm{\cpone}$ model}{The CP¹ model}}\label{cponemodel} 

We now turn to the prototypical theory in which holomorphic blocks
exhibit Stokes phenomena.  The $\cpone$ model can be
described in the ultraviolet as a gauged linear sigma model (GLSM)
\cite{Witten:1993yc} that flows, in the infrared, to a non-linear
sigma model with target space $\cpone$.  The GLSM is a U$(1)$ gauge
theory with two chiral fields $(\phi_1, \phi_2)$ that have same charges under the U$(1)$ gauge group.  The theory has a flavour
symmetry SU$(2)\times$U$(1)_J$ as well as a U$(1)_R$ symmetry\footnote{The SU(2)$\times$U(1) flavour symmetry might be enhanced to SU(3) in the IR as discussed recently in \cite{Gaiotto:2018yjh}.}.  The
flavour symmetry is broken to U$(1)_V\times$U$(1)_J$ by a twisted mass
$m$ for the fundamental flavours.  Similarly, we associate the FI
parameter $t$ to the U$(1)_J$ symmetry.  The scalar $\sigma$ in the 
vector multiplet is complexified by the Wilson line for the gauge field to a
field we denote $S$.  The twisted mass $m$ and FI parameter $t$ are similarly
complexified to $X$ and $Y$ by Wilson lines for the global U$(1)$ symmetries, respectively.  
As far as the 3d theory is concerned, the relevant variables and parameters are the exponentiated
ones that we denote as follows:
\be\label{eq:q-exact_CP1_v2:3}
  s = e^S\,,\qquad x = e^X\,, \qquad y = e^Y\,.
\ee
We assign the following charges and Chern-Simons coefficients \cite{Beem:2012mb}:
\be
T^{\cpone}[\vec{\phi}]=\left\{\quad
\begin{array}{c|cc}Q & \phi_1 & \phi_2 \\ \hline
G & 1 & 1 \\
V & 1 & -1 \\
J & 0 & 0 \\
R & 0 & 0
\end{array} \qquad
\begin{array}{c|cccc}k & G & V & J & R \\ \hline
G & 0 & 0 & 1 & 0 \\
V & 0 & 0 & 0 & 0 \\
J & 1 & 0 & 0 & 0 \\
R & 0 & 0 & 0 & \star
\end{array} \right.
\ee
Given this data, there is a systematic procedure described in 
\cite{Beem:2012mb} to derive an integral representation for the
holomorphic blocks and the line operator identities satisfied by the
block.  Since this is well-known, we simply state the result for the LOIs satisfied by the holomorphic blocks:
\begin{align}
\wh{p}_y +(\wh{y}^{-1} -\wh{x} -\wh{x}^{-1})+\wh{p}_y^{-1} &\simeq 0\,, \label{LOIcp11}\\
q^{-\frac{1}{2}}\wh{p}_x\wh{p}_y -\wh{x}\big(q^{\frac{1}{2}}\wh{p}_x +\wh{p}_y\big) +1 &\simeq 0\label{LOIcp12}\,.
\end{align}
Here we have defined the operators $\wh{p}_x = \sigma_q(x)$ and $\wh{p}_y= \sigma_q(y)$, which satisfy the $q$-commutation relations $\wh{p}_x \wh{x} =q\, \wh{x}\wh{p}_x$ and $\wh{p}_y \wh{y} =q\, \wh{y}\wh{p}_y$.

Let us first review the results of \cite{Beem:2012mb} in which the Stokes phenomenon is derived by making use of the
block-integral representation for the holomorphic blocks. This analysis is restricted to the mirror symmetry invariant plane in the complex $(X,Y)$ parameter space, given by $\text{Im}(X)=-\frac{2\pi}{3}$ and $\text{Im}(Y)=0$.
There are three Stokes regions and in each region, there are two solutions $(\bB^1,\bB^2)$ to the LOIs, which are associated to
particular contours in the block-integral representation. On
analytically continuing the parameters across Stokes lines, the blocks transform 
as shown in the Figure \ref{BDPStokesjumpsWblocks}.
\fig{!h}{\includegraphics[scale=1]{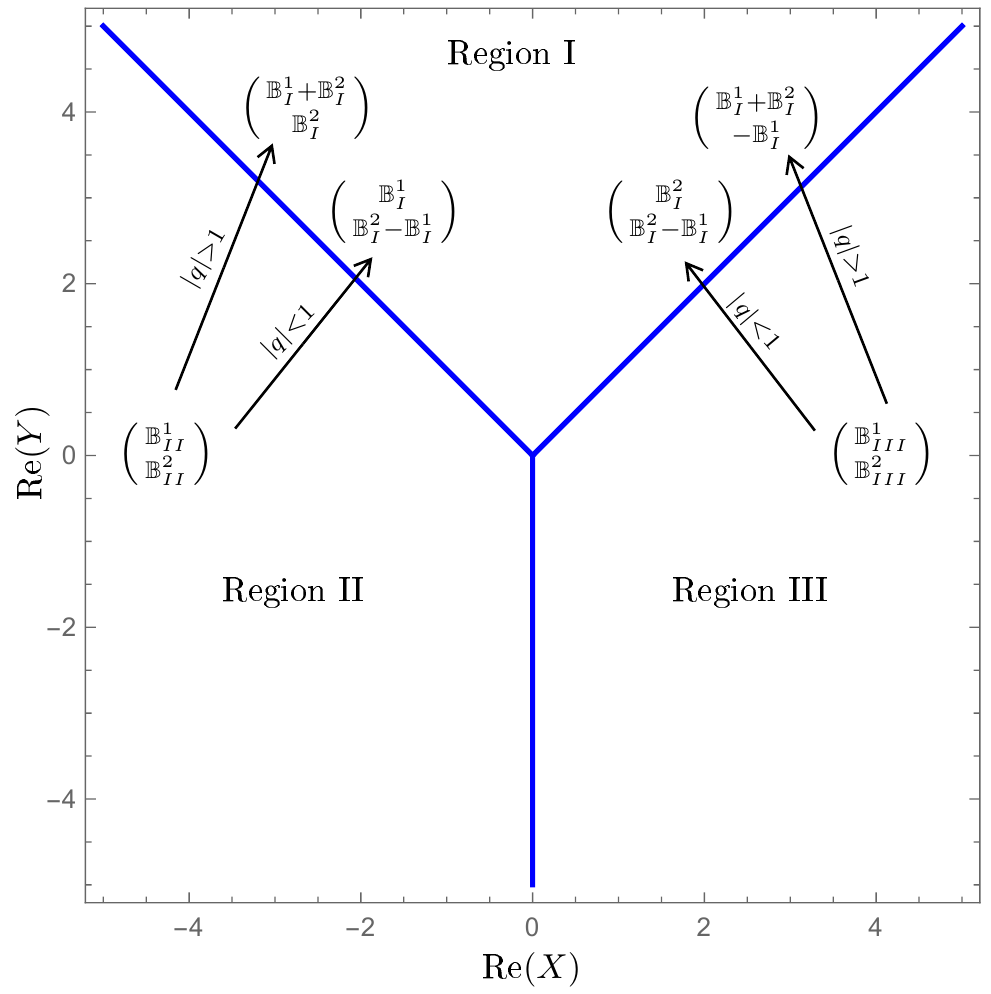}
\caption{The three Stokes regions for $\cpone$ at $\text{Im}(Y)=0$ and $\text{Im}(X)=-\frac{2\pi}{3}\cdot$ The analytic continuation of the blocks is shown from Regions II and III to Region I.}
\label{BDPStokesjumpsWblocks}
}

In what follows, we shall rederive all these results from a purely algebraic perspective.

\subsection{Holomorphic blocks}\label{sec:FirstHB}

We warm up by solving the LOIs in terms of $q$-hypergeometric
functions in a region of parameter space that we will eventually identify
as Region I in Figure \ref{BDPStokesjumpsWblocks}.  We shall focus first on
the LOI \eqref{LOIcp11} involving only $\wh{p}_y$. This LOI is a second order $q$-difference equation so has two independent solutions. Also, since this LOI is insensitive to purely $x$-dependent factors, we can write each block $\mathbb{B}$ as follows: 
\be
\label{basicblock1}
\mathbb{B}(x,y ;q) = f(x; q)\, g(x, y; q)\,,
\ee
where the function $f$ will be determined by solving the second LOI \eqref{LOIcp12}. We now define a new variable
\be
t = \frac{y}{qx} \,\cdot
\ee
In terms of this variable, we have $\wh{p}_y
\equiv\sigma_q(y)=\sigma_q(t)$. The LOI \eqref{LOIcp11} then takes the following form:
\be
\big[\sigma_q(t) +(qx)^{-1}t^{-1} -x -x^{-1} +\sigma_q^{-1}(t)\big]
g(x, t; q)=0\,.
\label{loi2}
\ee

In order to map this difference equation to the standard $q$-Goursat form\footnote{For
  details of the $q$-Goursat equation, refer to Appendix
  \ref{app:qGoursat}.}, we define a new function $h$ as follows: 
\begin{align}
g(x, t; q) &=\frac{\Theta _q (qxt)}{\Theta _q (qt)}\,h(x, t; q)\,.
\label{gvsh}
\end{align}
The function $h$ satisfies an equation that can be cast in the $q$-Goursat form as follows:
\begin{equation}
\Big[t P_1\big(\sigma_q(t)\big)-Q_1 \big(\sigma_q(t)\big)\Big]h(x,t;q)=0\,,
\label{q1}
\end{equation}
where the two $q$-difference operators $P_1$ and $Q_1$ are given by
\begin{align}
P_1\big(\sigma_q(t)\big)&=1\,; \\ 
Q_1\big(\sigma_q(t)\big)& =\left(-\frac{\sigma_q(t)}{q}\right)^{-1}\big(1-\sigma_q(t)\big)\left(1-\frac{a}{q}\sigma_q(t)\right) \quad\text{ with }\quad a = qx^2\,. 
\end{align}

\subsubsection{First holomorphic block}

One of the solutions to the equation \eqref{q1} is given by the $q$-hypergeometric function ${}_1\phi_1$: 
\be\label{hinf1}
h_1(x,t;q) = {}_1\phi_1\left(0;a;q, t^{-1} \right) = {}_1\phi_1\left(0;qx^2;q, qxy^{-1} \right).
\ee
The solution of the first LOI \eqref{LOIcp11} can then be written as 
\begin{align}
\bB^1(x, y; q)=f_1(x;q)\frac{\Theta_q(qxt)}{\Theta_q(qt)}\,{}_1\phi_1(0;qx^2;q,t^{-1}) =f_1(x;q)\frac{\Theta_q (y)}{\Theta_q (x^{-1}y)}{}_1\phi_1(0;qx^2;q,qxy^{-1})\,.
\end{align}
In order to solve for the block completely, we now fix $f_1(x; q)$ above by acting with the second LOI
\eqref{LOIcp12}.  The analysis is straightforward (but tedious) and we
end up with the following $q$-difference equation for $f_1$:
\be
f_1(qx;q)=\frac{q^{\frac{1}{2}}x}{(1-qx^2)(1-q^2 x^2)}f_1(x;q) \,,
\ee
which is solved by
\be
f_1(x)=\frac{(qx^2,q;q)_\infty}{\Theta _q (-\sqrt{q}x)} \,\cdot
\ee
So, the holomorphic block is given by
\begin{align}
\mathbb{B}^1(x,y;q) &= \frac{(qx^2 ,q;q)_\infty}{\Theta_q\left(-\sqrt{q}x\right)} \, \frac{\Theta_q(y)}{\Theta_q(x^{-1}y)}{}_1\phi_1(0;qx^2;q,qxy^{-1}) \nonumber\\
&= \frac{\Theta_q (y)}{\Theta _q \left(-\sqrt{q}x\right)\Theta_q(x^{-1}y)}\, {\mathcal J}( xy^{-1}, x^2; q)\,.
\label{block1R1} 
\end{align}
Here we have introduced the Hahn-Exton $q$-Bessel function denoted by ${\mathcal J}(x,y ;q)$ and whose properties are given in the Appendix \ref{specialfunctions}. In principle, the holomorphic blocks can be multiplied by an elliptic factor $E(x,y;q)$ that satisfies $\wh{p}_x E(x,y;q) = \wh{p}_y E(x,y;q) = E(x,y;q)$, since the modified blocks would also satisfy the same LOIs. We have chosen the above block to match the result of \cite{Beem:2012mb}. 

\subsubsection{Second holomorphic block}

In order to find the second solution, we have to analyze the singularity
structure of the $q$-difference equation \eqref{q1} for $h(t)$,
which can be written in the following form\footnote{We suppress the
arguments $\{x;q\}$ in $h(x,t;q)$ to avoid clutter and simplify expressions.}: 
\begin{equation}
[a\, t\sigma _q^{-1}(t)+ \big(1-(a+q)t \big)+qt\sigma_q(t)]h(t)=0\,.
\label{LOIht}
\end{equation}
We apply $\sigma _q (t)$ on the above
equation and obtain the following second order equation: 
\be
\sigma_q^2(t) h(t)=-\left(q^{-2}t^{-1}(1-a\, qt-q^2 t)\sigma_q(t) +q^{-1}a\right)h(t)\,.
\ee
Defining the two-component vector 
\begin{equation}
\Phi (t)=\begin{pmatrix}
h(t)\\
\sigma _q(t) h(t)
\end{pmatrix},
\end{equation}
the second order $q$-difference equation can be written as a matrix
equation of the form:
\be
\sigma _q(t) \Phi (t)=A(t)\Phi (t)\qquad \text{with}\qquad A(t)=
\begin{pmatrix}
0& 1\\
-\mu(t) & \lambda(t) 
\end{pmatrix},
\ee
where 
\be
\mu(t) = \frac{a}{q} \quad\text{and}\quad \lambda(t) =
\left(1+\frac{a}{q} - \frac{1}{q^2 t}\right).
\ee
From the coefficient matrix $A(t)$ we see that  $t \rightarrow \infty$
is a regular singular point and $t=0$ is an irregular singular
point.  We have already found one of the regular solutions near
$t=\infty$ in \eqref{hinf1}.  In order to find the other holomorphic
solution near $t=\infty$, let us write the coefficient matrix near
infinity:
\be
A(\infty)  =\begin{pmatrix}
0 & 1\\
-\frac{a}{q} & 1+\frac{a}{q}
\end{pmatrix}.
\ee
The eigenvalues of this matrix are $\{1, \frac{a}{q}\}$.  We will
restrict ourselves to the non-resonant case in which $a\ne q$ (which
is equivalent to the condition $x\ne \pm1$).  The procedure to obtain the
second solution is now standard (see \cite{TAB} for a review).  First
we write the character matrix
\be
\chi_{\infty}(t) = \begin{pmatrix}
1 & 0\\
0 & e_{\frac{a}{q}}(t) 
\end{pmatrix},\quad \text{ where }\quad e_{\omega}(t) =
\frac{\Theta_q(t)}{\Theta_q({\omega} t)}\,\cdot
\ee
For the case ${\omega}= \frac{a}{q}$, one can simplify this and
rewrite it as
\be
e_{\frac{a}{q}}(t) = \frac{q}{a} \, \frac{\Theta_q(qt)}{\Theta_q(a t) } \,\cdot
\ee
The second solution  is then written as 
\be
h_2(t) = e_{\frac{a}{q}}(t) \, \ell_2(t) \,,
\ee
where $\ell_2$ can be obtained by solving the matrix eigenvalue
equation: 
\be
\begin{pmatrix}
\ell_1(qt) & \ell_2(qt)\\
\ell_3(qt) & \ell_4(qt)
\end{pmatrix} 
\cdot
\begin{pmatrix}
1 & 0\\
0& \frac{a}{q}
\end{pmatrix} = \begin{pmatrix}
0 & 1\\
-\mu(t) & \lambda(t) 
\end{pmatrix}
\cdot
\begin{pmatrix}
\ell_1(t) & \ell_2(t)\\
\ell_3(t) & \ell_4(t)
\end{pmatrix}.
\ee
If we set $\ell_1(t) \equiv h_1(t)$, we
find that $\ell _2(t)$ satisfies the following
second order $q$-difference equation:
\be
\left( \sigma_q^2(t) - \frac{q}{a}\lambda(t) \sigma_q(t) +
  \frac{q^2}{a^2} \mu(t) \right)\ell_2(t) = 0\,.
\ee
One can map this to the standard $q$-Goursat form by the change of variables:
\be
\widetilde{t} =\frac{ a}{q}t\,, \qquad  \widetilde{a} = \frac{q^2}{a} \,\cdot
\ee
The second solution $h_2(t)$ can then be written (up to constant
prefactors) as
\be
h_2(t)= \frac{\Theta_q(qt)}{\Theta_q(a t) }\, {}_1\phi_1\left(0; \frac{q^2}{a}; q, \frac{q}{a t} \right).
\ee
The relation between $g_2$ (which solves \eqref{loi2}) and $h_2$ is
the same as in equation \eqref{gvsh}.  So all that remains to obtain the second independent
holomorphic block is to fix the prefactor $f_2$ in \eqref{basicblock1}. This can be done by using the second LOI; the
analysis is similar to what was done for the first block and we simply
present the final result (up to elliptic factors):  
\begin{align}
\mathbb{B}^2(x, y; q) &=\frac{(qx^{-2},q;q)_\infty\,\Theta_q(y)}{\Theta_q(-\sqrt{q}x)\,\Theta_q(xy)}\,{}_1\phi_1(0;qx^{-2};q,qx^{-1}y^{-1}) \nonumber\\ 
&= \frac{\Theta_q(y)}{\Theta_q(-\sqrt{q}x)\,\Theta_q(xy)}\,{\mathcal J}(x^{-1}y^{-1}, x^{-2}; q)\,.
\label{block2R1}
\end{align}

\subsubsection{Finding a region in parameter space}\label{firstreg1}

Given the explicit expression for the blocks in terms of the ${\mathcal J}$-functions, one can make use of the series expansion in \eqref{Jexpansion} to understand the region of validity of these expressions for the blocks in either $q$-chamber. Essentially the convergence of the expansion requires that for ${\mathcal J}(x,y;q)$, we have $|x| < 1$. For the holomorphic blocks in \eqref{block1R1} and \eqref{block2R1}, we obtain:
\be
\label{regionI}
\text{Re}(Y) - \text{Re}(X) > 0 \qquad\text{and}\qquad \text{Re}(Y) + \text{Re}(X) > 0\,.
\ee
This precisely maps to Region I in Figure \ref{BDPStokesjumpsWblocks} and we
infer that the blocks we have constructed are a basis of solutions to
the LOIs in Region I.  Henceforth, we shall denote these blocks as
$(\mathbb{B}^1_I, \mathbb{B}^2_I)$.  It is important to note here that
the blocks we have obtained in Region I in terms of $q$-hypergeometric
functions are valid expressions independent of whether $|q|<1$ or
$|q|>1$, since the method of solving the $q$-difference equations made no
assumptions regarding the $q$-chamber.

\section{\texorpdfstring{Stokes phenomena for 
    $\bm{\cpone}$ blocks}{Stokes phenomena for 
    CP¹ blocks}}\label{qstokesCP1} 

The holomorphic blocks we solved for in Region I are solutions to two $q$-difference equations.
Apart from meromorphic factors, the
non-trivial part is the $q$-hypergeometric function ${}_1\phi_1$ with
vanishing first argument.  Both of these solutions are analytic near
$t=\infty$. In this section, we compute non-trivial solutions of the $q$-difference equation satisfied by ${}_1\phi_1$ but near the
irregular singular point $t=0$ using the $q$-Borel and the $q$-Laplace
transformations discussed in Section \ref{qborel}. 

In particular, we derive in detail the connection formulae that relate
the solutions near the irregular singular point $t=0$ to those we
obtained near $t=\infty$.
Our claim is that these connection formulae fully encode and explain the Stokes
phenomena observed for the holomorphic blocks, which were derived
using the block-integral representation in \cite{Beem:2012mb}.
More importantly as we show, the same connection formula can contain
information about the analytic continuation to multiple regions in
parameter space for the $\cpone$ model.

We start with the equation \eqref{LOIht}:
\be
\left[at\sigma_q^{-1}(t) +(1-(a+q)t) +qt\sigma_q(t)\right] h(t)=0\,.
\ee
We look for solutions around $t=0$, and we assume the following ansatz
for the two linearly independent solutions \cite{RSZ, OHY}: 
\begin{align}
h_1(t) &= \frac{1}{\Theta _q (aqt)}\, u_1(t)\,, \label{h1ansatz} \\
h_2(t) &=\Theta_q(qt)\, u_2(t)\,. \label{h2ansatz}
\end{align}
The $q$-Borel resummation, as we shall see, will be done on the $u_k(t)$
part of the solution. We shall deal with each of these solutions in
turn, and we will see that the way $q$-Borel resummation is carried
out on the solution depends crucially on the chamber, i.e., whether
$|q| <1$ or $|q| >1$.

\subsection{Connection formulae for the first solution}\label{sec:CFfsol}

We begin with $h_1(t)$ and given the ansatz in \eqref{h1ansatz},
$u_1(t)$ satisfies the following equation
\be
\left[-aq^2 t^2\sigma _q(t) +(1-(a+q)t) -\sigma_q^{-1}(t)\right]u_1(t)=0 \,.
\ee
This can be solved by a power series solution of the form $u_1(t)=\sum
_{m\geq 0}a_m t^m$.  Taking the $q$-Borel transform of this equation
by ${\mathcal B}_q^-$ we find
\be
\left[aq^3\tau ^2 +(aq+q^2)\tau +1 -\sigma_q(\tau)\right]\mathcal{B}_q
^{-}[u_1](\tau)=0 \,, 
\ee
which is solved by
\be
\label{Bminusonh1}
\mathcal{B}_q ^{-}[u_1](\tau) =\frac{1}{(-q^2 \tau ,-aq\tau ;q)_\infty}\,\cdot
\ee

\subsubsection[\texorpdfstring{The $|q|<1$ chamber}{The |q|<1 chamber}]{The $\bm{|q|<1}$ chamber}

The $q$-Laplace transform for $|q|<1$ is given by
\begin{equation}
\mathcal{L}_q^{-}\left[\mathcal{B}_q ^{-}[u_1](\tau)\right](t)
=\frac{1}{2\pi i}\oint _{\Gamma _{\epsilon}} \frac{d\tau}{\tau}
\frac{\Theta_q\left(-t \tau^{-1}\right)}{(-q^2\tau,-aq\tau;q)_\infty}\,\cdot
\label{qlaplacemin}
\end{equation}
This integral can be done by deforming the contour and summing up the
contributions from the simple poles of $\mathcal{B}_q^-[u_1](\tau)$
for $|q|<1$ along with a possible contribution from $\tau=\infty$.  We
restrict the calculation to this $q$-chamber because the theta
function does not contribute to poles.  As shown in
\cite{Morita:2011hx, OHY}, the contribution from infinity vanishes for
this integral and so the $q$-Laplace transform reduces to minus the sum
over the poles of the $q$-Pochhammers in \eqref{qlaplacemin}. 

There are two different sets of poles which are located at
\begin{align}
\tau ^{(1)}_r =-q^{-(r+2)} \quad\text{ and }\quad \tau ^{(2)}_r=-a^{-1}q^{-(r+1)} \qquad\text{ for } r=0,1,\cdots\,.
\end{align}
The residue for the first set of poles $\tau ^{(1)}_r$ is given by
\begin{align}
-\text{Res}_{\tau ^{(1)}_r}= \frac{\Theta
  _q(q^2t)}{\left(\frac{a}{q},q;q\right)_\infty} \frac{(-1)^r
  q^{\frac{r}{2}(r-1)}}{\left(\frac{q^2}{a},q;q\right)_r}
  \left(\frac{q}{at}\right)^r.
\end{align} 
Denoting the sum over this first set of residues to be $I_1$, we obtain
\begin{align}
I_1=\frac{\Theta _q(q^2t)}{\left(\frac{a}{q},q;q\right)_\infty} {}_1
  \phi _1 \left( 0;\frac{q^2}{a};q,\frac{q}{at}\right).
\end{align}
Similarly, the residue from the other set of poles $\tau ^{(2)}_r$ is
given by
\begin{align}
-\text{Res}_{\tau ^{(1)}_r}=\frac{\Theta
  _q(aqt)}{\left(a,q;q\right)_\infty}\frac{(-1)^r
  q^{\frac{r}{2}(r-1)}}{\left(a,q;q\right)_r}
  \left(\frac{1}{t}\right)^r.
\end{align}
Denoting the sum over the second set of residues to be $I_2$, we
obtain
\begin{align}
I_2=\frac{\Theta _q(aqt)}{\left(\frac{q}{a},q;q\right)_\infty} {}_1
  \phi _1 \left( 0;a;q,\frac{1}{t}\right).
\end{align}
Thus, the connection formula in the $|q|<1$ chamber is given by
\begin{align}
h_1(t) &\longrightarrow \frac{1}{(qa^{-1},q;q)_\infty}{}_1 \phi_1\left(0;a;q,\frac{1}{t}\right) +
  \frac{1}{(aq^{-1},q;q)_\infty}\frac{\Theta _q(q^2 t)}{\Theta_q(aqt)}{}_1 \phi_1\left(0;\frac{q^2}{a};q,\frac{q}{at}\right) \nonumber\\
  &\longrightarrow \frac{1}{\Theta _q(a)}\mathcal{J}\left( q^{-1}t^{-1},q^{-1}a;q\right) +
  \frac{\Theta _q(q^2 t)}{\Theta _q(aq^{-1})\Theta_q(aqt)}\mathcal{J}\left( a^{-1}t^{-1},qa^{-1};q\right).
\label{connform1}
\end{align}
Here we have multiplied the $q$-Borel-resummed functions by the $\Theta_q$
prefactor in \eqref{h1ansatz}.  We recognize the $q$-hypergeometric
functions to be precisely those that appear in the holomorphic blocks
in Region I.  We now multiply the r.h.s by the following prefactor in
order to obtain the holomorphic blocks of the $\cpone$ theory:
\be
\label{Omega1defn}
\Omega_1(x,y;q) = -\frac{\Theta _q(qx^2)\Theta _q(y)}{\Theta
  _q(-\sqrt{q}x)\Theta _q(x^{-1}y)}\,\cdot
\ee
This factor is nothing but $f(x;q)\frac{\Theta_q(qxt)}{\Theta_q(qt)}$ arising due to \eqref{basicblock1} and \eqref{gvsh} and is essential for the block to satisfy both the LOIs of $\cpone$. By substituting $t^{-1}=qxy^{-1}$ and $a=qx^2$ in \eqref{connform1} and after some algebraic manipulations we find the following
connection formula: 
\begin{align}
\Omega_1 h_1 &\longrightarrow \mathbb{B}^2 _I -\mathbb{B}^1 _I\,.
\label{ar5} 
\end{align}
The above connection formula holds for $|q|<1$. The function appearing
on the l.h.s. will eventually be identified with the holomorphic block
in a Stokes region distinct from the Region I we have already
encountered; but in order to complete the identification, we also need to find the
connection formula in the $|q| > 1$ chamber.

\subsubsection[\texorpdfstring{The $|q|>1$ chamber}{The |q|>1 chamber}]{The $\bm{|q|>1}$ chamber}\label{newconfrm}

For $|q|>1$, the $q$-Laplace transform ${\mathcal L}^-_{q, \lambda}$ fails
to lead to a convergent integral. A similar problem for ${\mathcal L}^+_{q,\lambda}$ is encountered in the $|q|<1$ chamber in \cite{OHY}. The method proposed to deal with this issue was to introduce $p = \sqrt{q}$ and to perform the $p$-Borel transform followed by a $p$-Laplace transform. We have suitably adapted their methods for the $|q|>1$ case and have obtained the following connection formula (with $q=p^2$):
\begin{align}
\label{connformqmore1}
h_1(t)\longrightarrow \psi _1(t,p,\lambda)\, \mathcal{J}
  \left(a^{-1}t^{-1},q a^{-1};q\right) +\psi _2(t,p, \lambda)\,
  \mathcal{J}(q^{-1}t^{-1},q^{-1}a;q)\,, 
\end{align}
where the coefficients $\psi_i$ are given by
\begin{align}
\psi _1(t,p,\lambda) &=\frac{\Theta _{q}(apq^2 \lambda ^2)\Theta _{q} (-p\lambda t^{-1})\Theta _{q} (-q\lambda t^{-1})}{\Theta _{q}(aq t)\Theta _{q}(-ap\lambda)\Theta _{q}(-ap^2 \lambda)\Theta _{q}(\lambda^2 p^5 t^{-1})}\,, \\
\psi _2(t,p,\lambda) &=\frac{\Theta _{q}(a^{-1}p^{-3}\lambda ^{-2})\Theta _{q}(-p\lambda t^{-1})\Theta _{q}(-q\lambda t^{-1})}{\Theta _{q}(aq t)\Theta _{q}(-q^{-1}\lambda^{-1})\Theta _{q}(-p^{-1}\lambda^{-1})\Theta _{q}(apq\lambda^2 t^{-1})}\,\cdot
\end{align}
This is the main new result of this work and it is proved in detail in
Appendix \ref{proofofconn}.  According to \cite{OHY,DRELO} and as discussed in subsection \ref{sec:qLaplaceT}, $\lambda$ is such that $\left[\lambda\right] \in \left(\mathbb{C}^{\star}/q^\mathbb{Z}\right)\setminus \{\left[1\right] \}$. We now
claim that the holomorphic blocks (built on the $h_1(t)$) satisfy both the LOIs of $\cpone$
model only for two values of $\lambda$. Furthermore, we find that precisely for these values, the
connection formula in \eqref{connformqmore1} coincides with those
derived in \cite{Beem:2012mb} by using the block-integral
representation.  Let us see this in detail.
\begin{enumerate}
\item We set $\lambda =-p^{-1}=-q^{-\frac{1}{2}}$ and substitute
  $t=\frac{y}{qx}$ and $a=qx^2$. The coefficients $\psi_i$ then take the
  following simplified values:
\begin{align}
\psi _1 =-\frac{\Theta _q(x^{-1}y)}{\Theta _q(qx^2)\Theta _q(xy)}\,;\qquad \psi _2 =0\,. 
\end{align}
The vanishing of $\psi _2$ is due to the factor $\Theta
_q(-p^{-1}\lambda ^{-1})$ in the denominator, which tends to infinity for the choice of
$\lambda =-p^{-1}$.  The connection formula therefore simplifies to
the following form: 
\begin{align}
h_1(t)\longrightarrow-\frac{\Theta _q(x^{-1}y)}{\Theta _q(qx^2)\Theta _q(xy)}\mathcal{J}(x^{-1}y^{-1},x^{-2};q)\,.
\end{align}
Now to obtain the holomorphic block of the $\cpone$ theory we
multiply with the same factor $\Omega_1$ as in \eqref{Omega1defn}.
We thereby obtain the following connection formula for the solution
around $t=0$ for this particular value of $\lambda$:
\begin{align}
\Omega_1 h_1 \longrightarrow &\; \frac{\Theta _q(y)}{\Theta _q(-\sqrt{q}x)\Theta _q(xy)} \mathcal{J}(x^{-1}y^{-1},x^{-2};q) \nonumber\\
=&\; \mathbb{B}^2_I\,.
\end{align}

\item We set $\lambda =-pa^{-1}=-q^{-\frac{1}{2}}x^{-2}$ and the
  coefficients $\psi_i$ take the following values:
\begin{align}
\psi _1 =0\,;\qquad \psi _2 =\frac{1}{\Theta _q (qx^2)}\,\cdot
\end{align}
The vanishing of $\psi_1$ is because $\Theta _q(-ap \lambda)$ tends to infinity
for $\lambda = -q^{-\frac{1}{2}}x^{-2}$.  Using the same
multiplicative factor $\Omega_1$ to satisfy the LOIs of the
$\cpone$ theory we obtain the analytically continued solution: 
\begin{align}
\Omega_1 h_1 \longrightarrow & -\frac{\Theta _q(y)}{\Theta _q(-\sqrt{q}x)\Theta _q(x^{-1}y)} \mathcal{J}(xy^{-1},x^2;q)\nonumber\\
=&-\mathbb{B}^1_I\,.
\end{align}

\end{enumerate} 

We summarize the results of this subsection in Figure \ref{Omega1h1conn}.
\fig{!h}{\includegraphics[scale=1]{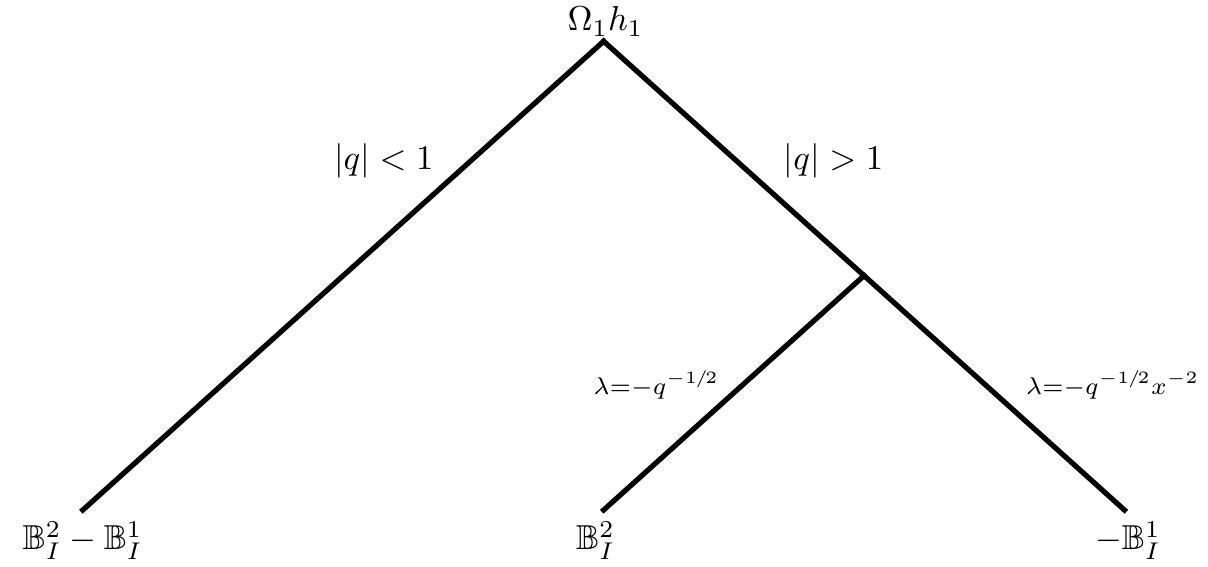}
\caption{Summary of the connection formula for the solution
  $\Omega_1h_1$. All connection formulae are obtained by first using
  the ${\mathcal B}^-$ Borel transform, followed by the inverse Laplace
  transform ${\mathcal L}^-$. For $|q| <1$ there is a unique way to
  analytically continue the solution. For $|q| >1$, there are two
  ways to analytically continue the solution consistent with both LOIs of the $\cpone$ theory, as shown.}
\label{Omega1h1conn}
}

\subsection{Connection formulae for the second solution}

We now turn to the second solution $h_2(t)=\Theta_q(qt)u_2(t)$. The $q$-difference equation satisfied by $u_2(t)$ is given by
\begin{align}
\left[ -\sigma_q(t) + (1-(a+q)t) - a\, t^2\, \sigma_q^{-1}(t)\right] u_2(t) = 0\,.
\end{align}
We act with the $q$-Borel transform ${\mathcal B}^+_q$ on the above equation and using the identities in \eqref{Bplusminusids}, one can check that the $q$-Borel transformed solution satisfies the following
equation:
\begin{align}
\sigma _q(\tau) \mathcal{B}_q^{+}[u_2](\tau)= \frac{1}{(a\tau +1)(q\tau +1)}\mathcal{B}_q^{+}[u_2](\tau)\,,
\end{align}
which is solved by 
\begin{equation}
\mathcal{B}_q^{+}[u_2](\tau) =\left(-a\tau ,-q\tau;q\right)_\infty \,.
\end{equation}

\subsubsection[\texorpdfstring{The $|q|>1$ chamber}{The |q|>1 chamber}]{The $\bm{|q|>1}$ chamber}

For $|q|>1$, the $q$-Pochhammer has simple poles and as we did for the
previous solution, it is possible to use the $q$-Laplace transform
${\mathcal L}_q^+$ in equation \eqref{qlaplace1} to obtain the $q$-Borel
resummed solution:
\begin{equation}
\label{LplusonBplusone}
\mathcal{L}_q ^{+} \left[\mathcal{B}_q^{+}[u_2](\tau)\right](t) =\frac{1}{2\pi i} \oint _{\Gamma_\epsilon} \frac{d\tau}{\tau}\left(-a\tau ,-q\tau;q\right)_\infty \Theta _q^{-1}\left(-q t \tau^{-1}\right),
\end{equation}
where $\Gamma _\epsilon$ is a contour that encircles the origin.  There are two infinite sets of poles, located at 
\be
\tau^{(1)}_r =-a^{-1}q^{r+1} \quad\text{ and }\quad \tau^{(2)}_r =-q^{r} \qquad\text{ for } r=0,1, \cdots\,.
\ee
Then the residue for the first set of poles $\tau^{(1)}_r$ is given by
\begin{equation}
-\text{Res}_{\tau^{(1)}_r} =\frac{(q^2a^{-1};q)_\infty (q^{-1};q^{-1})_\infty}{\Theta _q(t a)} \frac{(-1)^r q^{\frac{r}{2}(r-1)}(qt ^{-1} a^{-1})^r}{(q^2a^{-1},q;q)_r}\,\cdot
\end{equation}
Then we can analytically do the complex integral for $|q|>1$ by
summing up all the residues for $r \in [0,\infty]$,  which gives 
\begin{align}
I_1 =\frac{(q^2a^{-1};q)_\infty (q^{-1};q^{-1})_\infty}{\Theta _q(t
  a)} ~{}_1 \phi _1(0;q^2a^{-1};q,qt ^{-1}a^{-1})\,.
\end{align}
The sign is reversed due to the fact that we have to consider minus the sum of residues at the poles $\tau^{(1)}_r$. Similarly, the residues arising from the second set of poles $\tau^{(2)}_r$ are given by
\begin{equation}
-\text{Res}_{\tau^{(2)}_r}
=\frac{(a;q)_\infty (q^{-1};q^{-1})_\infty}{\Theta _q(qt)}
  \frac{(-1)^n q^{\frac{n}{2}(n-1)}t ^{-n}}{(a,q;q)_n}\,\cdot
\end{equation}
Summing over all residues, we obtain 
\begin{align}
I_2 =\frac{(a;q)_\infty (q^{-1};q^{-1})_\infty}{\Theta _q(qt)}  ~{}_1
  \phi _1 (0;a;q,t ^{-1})\,.
\end{align}
It is interesting to observe that we obtain the same $q$-hypergeometric series as we did in $|q|<1$ chamber.  Putting together
the two contributions $I_1$ and $I_2$, and rewriting the
$q$-hypergeometric functions in terms of the ${\mathcal J}$-function,
we obtain the connection formula for $|q|>1$:
\begin{equation}
\label{h2conn}
h_2(t) \longrightarrow (q^{-1};q^{-1})_\infty \left[\mathcal{J}(q^{-1}t^{-1},q^{-1}a;q) +\frac{\Theta _q(qt)}{\Theta _q(at)} \mathcal{J}(t^{-1}a^{-1},qa^{-1};q) \right].
\end{equation}
Now to apply this connection formula to the blocks in the
$\cpone$ case we have to multiply by the following prefactor (which follows from an analysis similar to the one that led to $\Omega_1$):
\be\label{Omega2defn}
\Omega_2(x,y;q) = \frac{\Theta _q(y)}{\Theta _q(-\sqrt{q}x)\Theta _q(x^{-1}y)}\,\cdot
\ee
By setting $t^{-1}=qxy^{-1}$ and $a=qx^2$, and identifying the terms
on the r.h.s of the connection formula in \eqref{h2conn} with the
holomorphic blocks in Region I of the $\cpone$ theory, we obtain the
following connection formula for $|q|>1$:
\begin{align}
\Omega_2 h_2 \longrightarrow \mathbb{B}^1_I +\mathbb{B}^2_I \,.
\end{align} 

\subsubsection[\texorpdfstring{The $|q|<1$ chamber}{The |q|<1 chamber}]{The $\bm{|q|<1}$ chamber}

The connection formula for this case has been derived in \cite{OHY}
and we simply present the result relating the solution near $t=0$ to the
regular solutions near $t=\infty$:
\be
h_2(t) \longrightarrow \chi _1(t,p,\lambda)~ \mathcal{J}
  \left(q^{-1}t^{-1},q^{-1}a;q\right) +\chi _2(t,p,\lambda)~ \mathcal{J}
  \left(a^{-1}t^{-1},qa^{-1};q\right),
\ee
where the coefficients $\chi_i$ are given by
\begin{align}
\chi _1(t,p,\lambda) &=\frac{\Theta _q (qt)\Theta _q (-a\lambda)\Theta_q (-a\lambda p)\Theta _q \left(pt\lambda^{-2}\right)}{\Theta _q (ap \lambda^2)\Theta _q \left(-pt\lambda^{-1}\right)\Theta_q \left(-qt\lambda^{-1}\right) \Theta _q(qa^{-1})}\,, \label{plusconn1}\\
\chi _2(t,p,\lambda)&=\frac{\Theta _q (qt)\Theta _q (-q\lambda)\Theta _q (-pq\lambda )\Theta _q \left(pqt a^{-1}\lambda^{-2}\right)}{\Theta _q \left(-pt\lambda^{-1}\right)\Theta _q \left(-qt\lambda^{-1}\right)\Theta_q(ap\lambda^2) \Theta _q(aq^{-1})} \label{plusconn2}\,\cdot
\end{align}
As for the case studied previously, the connection formula involves
the parameter $\lambda$ and we choose the same two values as
before.  Exactly for these values, it turns out that these provide
solutions to both the LOIs of the $\cpone$ theory.

\begin{enumerate}
\item We choose the parameter $\lambda = -p^{-1}=-q^{-\frac{1}{2}}$, for
  which the coefficients become
\begin{align}
\chi _1=\frac{\Theta _q (a)}{\Theta_q \left(qa^{-1}\right)}=1\,;\qquad \chi _2=0\,.
\end{align}
The last equality is due to the factor $\Theta _q (-pq\lambda)=\Theta _q (q )$, which vanishes.
To satisfy both the LOIs, we have to multiply the above with the same
prefactor $\Omega_2$ given in \eqref{Omega2defn}. Substituting $t=\frac{y}{qx}$
and $a=qx^2$ we obtain the following connection formula in terms of the holomorphic blocks:
\begin{align}
\Omega_2 h_2 \longrightarrow&\; \frac{\Theta _q (y) }{\Theta _q (-\sqrt{q}x)\Theta _q(x^{-1}y)}\mathcal{J}(xy^{-1},x^2;q) \nonumber\\
=&\; \mathbb{B}^1 _I  \,.
\end{align}

\item The second consistent choice of $\lambda$ is given by $\lambda
  =-p a^{-1} =-q^{-\frac{1}{2}}x^{-2} $.  Due to the vanishing factor
  $\Theta _q (-a\lambda p)=\Theta _q (p^2) =0$, we find that
\begin{align}
\chi _1  =0\,; \qquad \chi _2 =\frac{\Theta _q (x^{-1}y)\Theta _q (qx^{-2})}{\Theta _q(xy)\Theta_q \left(x^2\right)} \,\cdot
\end{align}
The connection formula in terms of blocks then reads:
\begin{align}
\Omega _2 h_2 \longrightarrow &\; \frac{ \Theta _q (y)}{\Theta _q (-\sqrt{q}x)\Theta _q (xy)}\mathcal{J}(x^{-1}y^{-1},x^{-2};q) \nonumber \\
=&\;\mathbb{B}_I ^2 \,.
\end{align}

\end{enumerate}

We summarize the results of this subsection in Figure \ref{Omega2h2conn}.
\fig{!h}{\includegraphics[scale=1]{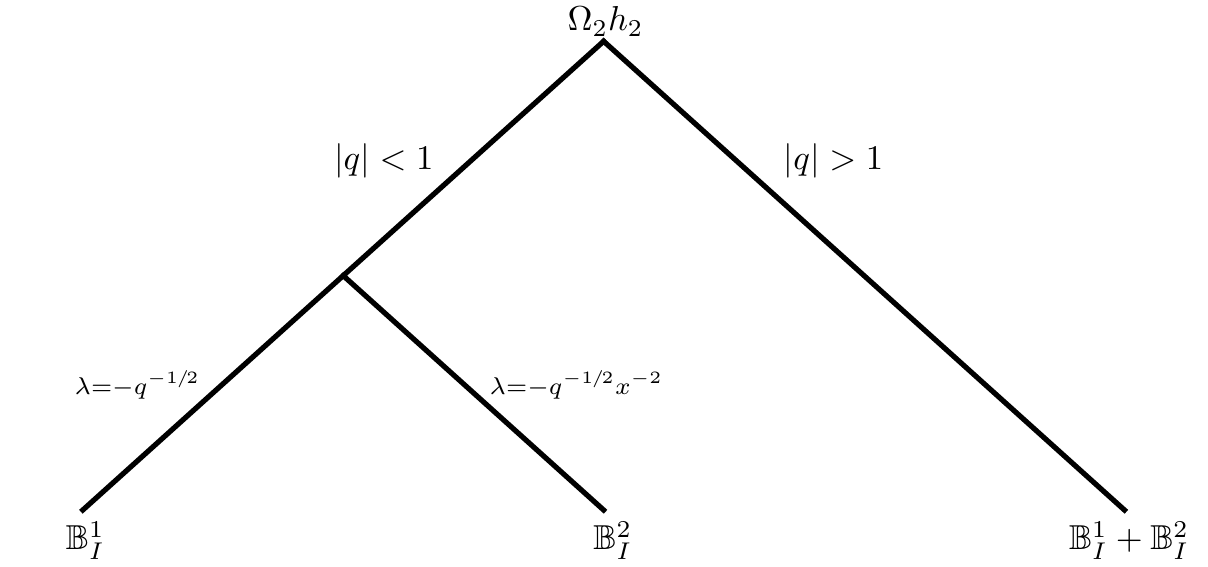}
\caption{Summary of the connection formula for the solution
  $\Omega_2h_2$. All connection formulae are obtained by first using
  the ${\mathcal B}^+$ Borel transform, followed by the inverse Laplace
  transform ${\mathcal L}^+$. For $|q| >1$ there is a unique way to
  analytically continue the solution. For $|q| <1$,  there are two
  ways to analytically continue the solution consistent with both LOIs of the $\cpone$ theory, as shown.}
\label{Omega2h2conn}
}

\section{Concluding remarks}\label{Stokesfromconn}

We have focussed on holomorphic blocks of the $\cpone$
model. Semi-classically, the theory has two massive vacua and from the
general analysis of \cite{Beem:2012mb}, it follows that there are 
two holomorphic blocks in any given region of parameter space labelled
by the complexified twisted mass and FI parameter $(X,Y)$, respectively. The
blocks, in turn, are solutions to two linear second order $q$-difference equations. In one
region of parameter space, we could easily solve both the LOIs in terms of the 
$q$-hypergeometric function ${}_1\phi_1$ as the first LOI \eqref{LOIcp11} has a
regular singular point. We denoted these blocks by $(\mathbb{B}_I^1,\mathbb{B}_I^2)$.

Obtaining the solutions in the other regions of parameter space turned
out to be more subtle because these correspond to the solutions of first LOI \eqref{LOIcp11}
near an irregular singular point. In the nomenclature of \cite{OHY} (see also
the talk \cite{OHYtalk} where more details are given), in each chamber
(either $|q|<1$ or $|q|>1$), one of the solutions is a convergent series while the other is a divergent series.
For the convergent series there is a unique way to analytically continue
the solution while the analytic continuation of the divergent series
depends on a complex parameter $\lambda$.  In other words, for a
given $q$-divergence series, the Stokes region depends on $\lambda$ \cite{OHYtalk}.  

Taking into account the second LOI \eqref{LOIcp12} satisfied by the holomorphic blocks of the $\cpone$ theory, it turned out that only for two values of $\lambda$ did the analytic continuation lead to consistent holomorphic blocks. Putting these different mathematical notions together, one expects three Stokes regions in the $\cpone$ theory, which agrees with what was found in \cite{Beem:2012mb}. In order to make more precise comparisons with their results, let us rearrange the connection formulae summarized in Figure \ref{Omega1h1conn} and Figure \ref{Omega2h2conn} such that we pair the blocks together for a given value of $\lambda$, as shown in Figure \ref{analyticnew}. We observe that we exactly reproduce the results of \cite{Beem:2012mb} shown in Figure \ref{BDPStokesjumpsWblocks} by noting that the right half of Figure \ref{analyticnew} corresponds to analytic continuation of blocks from Region II and the left half to that from Region III. It is thus the choice of $\lambda$ that effectively distinguishes the Stokes regions.
\fig{!h}{\includegraphics[scale=1]{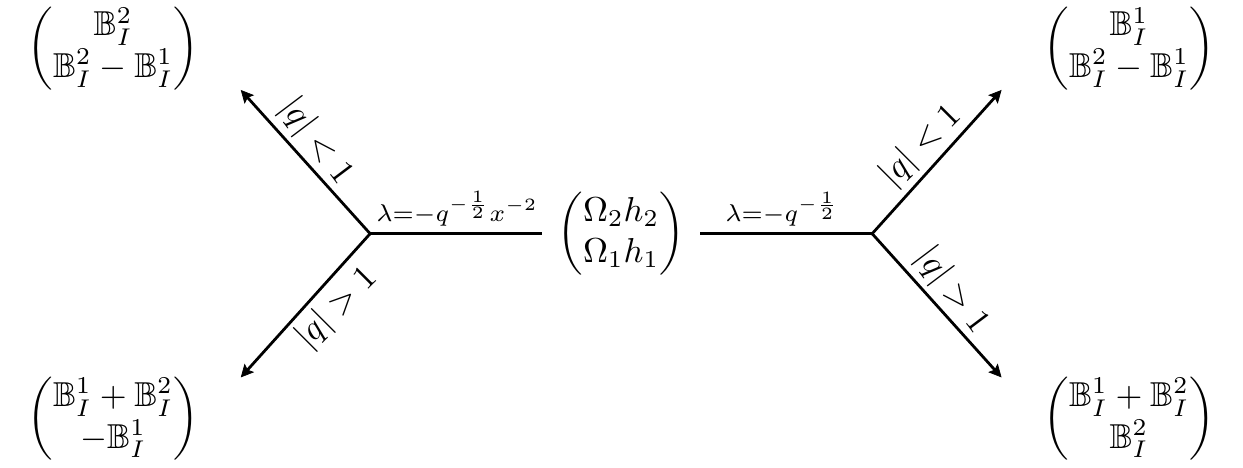}
\caption{Analytic continuation of the blocks for $\lambda = -q^{-\frac{1}{2}}$ and $\lambda = -q^{-\frac{1}{2}} x^{-2}$.}
\label{analyticnew}
}

It now remains to identify the precise expressions for the blocks in the different Stokes regions. While the path to do this has already been outlined in \cite{Beem:2012mb} we now comment on how the algebraic approach adds to the discussion. The basic idea in \cite{Beem:2012mb} is to exploit the fact that while the holomorphic blocks have different analytic behaviour in the $|q|<1$ and $|q| >1$ chambers, they have identical series expansions. So, if one is able to find an expression for the block in a given chamber as a single ${\mathcal J}$-function, the $q$-expansion in the other chamber is guaranteed to be the same. However, given such a form for the block in one chamber, the algebraic approach guarantees that it will have the correct analytic continuation in the other chamber. 

We begin with the $\lambda=-q^{-\frac{1}{2}}$ case and consider the  analytic continuation results for $\Omega_2 h_2$ in $|q|<1$ chamber and $\Omega_1 h_1$ in $|q|>1$ chamber (right half of Figure \ref{analyticnew}).  We see that these coincide with individual blocks valid in the Region I as derived in subsection \ref{sec:FirstHB}. Following \cite{Beem:2012mb} we use the identities \eqref{Jqlessone} and \eqref{Jqmoreone} valid in the respective $q$-chambers to rewrite the $\mathcal{J}$-functions and claim that the pair of blocks for $\lambda=-q^{-\frac{1}{2}}$ is given by
\begin{align}
 \mathbb{B}_{R_1}:=\Big(\Omega_2 h_2 \Big)_{\lambda=-q^{-\frac{1}{2}}}  &= \frac{\Theta _q (y)}{\Theta _q(-\sqrt{q}x)\Theta
_q(x^{-1}y)} \mathcal{J}(x^2,xy^{-1};q)\,. \\
 \mathbb{B}_{R_2}:=\Big(\Omega_1 h_1\Big)_{\lambda=-q^{-\frac{1}{2}}}  &= -\frac{\Theta _q(y)\Theta _q(qx^2)}{\Theta_q(-\sqrt{q}x)\Theta _q(qxy)\Theta _q(x^{-1}y)}\mathcal{J}(xy,x^{-1}y;q)\,. 
\end{align}
Note that we have used an identity valid only in one
$q$-chamber and not the other to write the above expressions. In \cite{Beem:2012mb} a formula for the
${\mathcal J}$-function was conjectured in order to make sense of the  
connection formula in the other chamber. From the perspective of blocks as contour integrals, it is not obvious why these ${\mathcal J}$-functions can be continued to give the results in the other $q$-chamber. However, from the explicit analytic continuation of the $\Omega_i h_i$ for this value of $\lambda$ derived in the previous section, it is guaranteed and can be taken as a proof of that conjectured formula.

A similar analysis can be done for the other value of $\lambda
=-q^{-\frac{1}{2}}x^{-2}$ by making use of the same identities 
\eqref{Jqlessone} and \eqref{Jqmoreone} but focussing on the
left half of Figure \ref{analyticnew}:
\begin{align}
\mathbb{B}_{R_3} &:=\Big(\Omega_2 h_2\Big)_{\lambda=-q^{-\frac{1}{2}}x^{-2}
                   }=\frac{\Theta _q (y)}{\Theta _q(-\sqrt{q}x)\Theta
  _q(xy)} \mathcal{J}(x^{-2},x^{-1}y^{-1};q)\,. \\
\mathbb{B}_{R_4} &:=\Big(\Omega_1 h_1\Big)_{\lambda=-q^{-\frac{1}{2}}x^{-2}
                   }=\frac{\Theta _q(y)\Theta _q(x^2)}{\Theta
   _q(-\sqrt{q}x)\Theta _q(xy)\Theta _q(xy^{-1})}
   \mathcal{J}(x^{-1}y,xy;q)\,.
\end{align}

We now claim that the constraints arising from the convergence of the
power series expansions of these new ${\mathcal J}$-functions should
lead to new Stokes regions, where these blocks form a well-defined
basis. Furthermore, we will show that the pairing based on the choice
of $\lambda$ suggested above and the convergence properties are 
consistent because the domains of validity of the three pairs of
blocks allow a single covering of the parameter space. The parameter
space that is under consideration  is spanned by $X$ and $Y$ defined
in \eqref{eq:q-exact_CP1_v2:3}. The blocks are written in terms of the
$\mathcal{J}$-functions up to rational products of theta factors.  Each
$\mathcal{J}(x,y;q)$ function is defined in the $|x|<1$ region. The
region in which a pair of blocks is defined, is determined by the
overlap of the domains of validity of the two $\mathcal{J}$-functions
associated to that pair of blocks. We work these regions out in detail
now.

\begin{figure}[!h]
\centering
\begin{tikzpicture}[scale=0.50]
\draw[blue, ultra thick] (-5.0,5.0) -- (0.0,0.0) ;
\draw[blue, ultra thick] (0.0,0.0) -- (0.0,-5.0) ;
\draw[blue, ultra thick] (5.0,5.0) -- (0.0,0.0) ;
\draw[black, dotted] (0.0, 0.0) -- (5.0, -5.0) ;
\draw[black, dotted] (0.0, 0.0) -- (0.0, 5.0) ;
\draw[black, dotted] (0.0, 0.0) -- (-5.0, -5.0) ;
\filldraw[black] (-1.5,4.5) circle (0pt) node[anchor=west]{Region I} ;
\filldraw[black] (-4.5,-1.5) circle (0pt) node[anchor=west]{Region II} ;
\filldraw[black] (1.0,-1.5) circle (0pt) node[anchor=west]{Region III} ;
\draw[black, ->] (4.0, 4.0) -- (3.5, 4.5) ;
\draw[black, ->] (-3.5, -3.5) -- (-3.8, -3.2) ;
\filldraw[black] (3.0,4.9) circle (0pt) node[anchor=west]{{\small $\mathbb{B}^{1}_I$}} ;
\filldraw[black] (-4.1,-2.9) circle (0pt) node[anchor=west]{{\tiny $\mathbb{B}^{1}_I$}} ;
\draw[black, ->] (-4.0, 4.0) -- (-3.5, 4.5) ;
\draw[black, ->] (3.5, -3.5) -- (3.8, -3.2) ;
\filldraw[black] (-4.0,4.9) circle (0pt) node[anchor=west]{{\small $\mathbb{B}^{2}_I$}} ;
\filldraw[black] (3.5,-2.9) circle (0pt) node[anchor=west]{{\tiny $\mathbb{B}^{2}_I$}} ;
\draw[black, ->] (0.0, -3.5) -- (-0.71, -3.5) ;
\draw[black, ->] (0.0, 3.5) -- (-0.42, 3.5) ;
\filldraw[black] (-2.2,-3.5) circle (0pt) node[anchor=west]{{\small $\mathbb{B}_{R_1}$}} ;
\filldraw[black] (-1.65,3.5) circle (0pt) node[anchor=west]{{\tiny $\mathbb{B}_{R_1}$}} ;
\draw[black, ->] (-3.5, 3.5) -- (-4.0, 3.0) ;
\draw[black, ->] (4.0, -4.0) -- (3.7, -4.3) ;
\filldraw[black] (-4.9,2.7) circle (0pt) node[anchor=west]{{\small $\mathbb{B}_{R_2}$}} ;
\filldraw[black] (3.1,-4.6) circle (0pt) node[anchor=west]{{\tiny $\mathbb{B}_{R_2}$}} ;
\draw[black, ->] (0.0, -4.21) -- (0.71, -4.21) ;
\draw[black, ->] (0.0, 2.79) -- (0.42, 2.79) ;
\filldraw[black] (0.55,-4.21) circle (0pt) node[anchor=west]{{\small $\mathbb{B}_{R_3}$}} ;
\filldraw[black] (0.25,2.79) circle (0pt) node[anchor=west]{{\tiny $\mathbb{B}_{R_3}$}} ;
\draw[black, ->] (3.5, 3.5) -- (4.0, 3.0) ;
\draw[black, ->] (-4.0, -4.0) -- (-3.7, -4.3) ;
\filldraw[black] (3.9,2.7) circle (0pt) node[anchor=west]{{\small $\mathbb{B}_{R_4}$}} ;
\filldraw[black] (-3.85,-4.6) circle (0pt) node[anchor=west]{{\tiny $\mathbb{B}_{R_4}$}} ;
\draw[black, ->] (5.5, 0.0) -- (5.5, 1.0) ;
\draw[black, ->] (5.5, 0.0) -- (7.75, 0.0) ;
\filldraw[black] (5.35,0.4) circle (0pt) node[anchor=west]{{\tiny \text{Re}(X,Y)}} ;
\end{tikzpicture}
\caption{Thick (blue) lines are the Stokes lines dividing the
  $\text{Re}(X,Y)$-plane into three Stokes regions. The domain of validity of
  each block is shown by arrows. The basis of blocks in each Stokes
  region is given by the two blocks in larger font. (Dotted lines play
  no role in defining the Stokes regions.)} 
\label{fig:contour}
\end{figure}
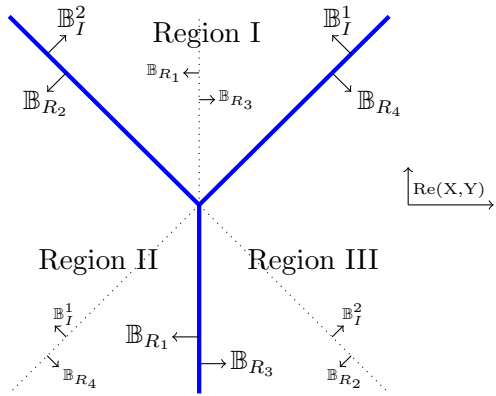

For the original pair of blocks, namely $\mathbb{B}^1_{I}$ and
$\mathbb{B}^2_{I}$, this was already done in the subsection \ref{firstreg1} and
the region where both the blocks are defined is given by the overlap of
\begin{equation}
  \label{eq:q-exact_CP1_v2:4}
  \text{Re}(X)-\text{Re}(Y)<0 \quad \text{ and }\quad \text{Re}(X)+\text{Re}(Y)>0\,.
\end{equation}
This overlap region, which we call Region I, is a right-angled wedge
in the upper half of $\text{Re}(X,Y)$-plane.  Let us apply the same
idea to the other two pairs of blocks.  The pair of blocks
($\mathbb{B}_{R_1}$, $\mathbb{B}_{R_2}$) are defined in the region
bounded by
\begin{equation}
  \label{eq:q-exact_CP1_v2:5}
  \text{Re}(X)<0 \quad \text{ and }\quad \text{Re}(X)+\text{Re}(Y)<0\,.
\end{equation}
This region (Region II) is bounded by the negative $\text{Re}(Y)$-axis
and the $\text{Re}(X)+\text{Re}(Y)=0$ line in the upper half of
$\text{Re}(X,Y)$-plane. Notice the latter boundary matches with the
second boundary ($\text{Re}(X)+\text{Re}(Y)=0$) of the Region
I. Finally, the pair of blocks ($\mathbb{B}_{R_3}$,
$\mathbb{B}_{R_4}$) are defined in the region (Region III) bounded by
\begin{equation}
  \label{eq:q-exact_CP1_v2:6}
  \text{Re}(X)>0 \quad \text{ and }\quad \text{Re}(X)-\text{Re}(Y)>0\,.
\end{equation}
It is easy to see that Region III is bounded by the negative
$\text{Re}(Y)$-axis and the line $\text{Re}(X)-\text{Re}(Y)=0$ in the
upper half of $\text{Re}(X,Y)$-plane.  The former boundary matches
with the first boundary ($\text{Re}(X)=0$) of the Region II and the
latter boundary matches with the first boundary
($\text{Re}(X)-\text{Re}(Y)=0$) of the Region I. This completes the
(re)derivation of the Stokes regions in the $\cpone$ model and
we summarize this discussion in the Figure \ref{fig:contour}.

It is useful to recall that in \cite{Beem:2012mb} these Stokes regions were obtained by using various means including the self-mirror property of the $\cpone$ model. In the algebraic approach, we did not use such physical considerations and focused solely on the LOIs and the analytic continuation of the $q$-hypergeometric functions. It would be interesting to see if this approach can be generalized to other models with irregular singular points for the blocks such as the $\mathbb{CP}^N$ models where the block-integrals and contour deformation methods might be more difficult to implement.

\acknowledgments We would like to thank Renjan R. John, Alok Laddha and Madhusudhan Raman for useful discussions. We especially thank Tudor Dimofte for his helpful correspondence regarding \cite{Beem:2012mb} and his useful comments on the draft of this work. SA would like to thank the \'Ecole Normale Sup\'erieure, Paris and the Universit\`a di Torino, Italy for their hospitality during the completion of this work. DJ would like to thank the Institute of Mathematical Sciences, Chennai for generous hospitality during the early stages of this work. AM would like to thank Harish-Chandra Research Institute, Allahabad for hospitality during the completion of this work. This research was supported in part by the International Centre for Theoretical Sciences (ICTS) during a visit for participating in the program - Quantum Fields, Geometry and Representation Theory (Code: ICTS/qftgrt/2018/07).

\begin{appendix}
\section{\texorpdfstring{The $\bm{q}$-Goursat equation}{The q-Goursat equation}}\label{app:qGoursat}

The $q$-analogue of Goursat's equation is given as follows (we follow the conventions in \cite{OHYtalk, Adachi}): 
\be
[x P_r\big(\sigma_q(x)\big)-Q_s\big(\sigma_q(x)\big)]f(x)=0\,,
\label{qrs}
\ee
where the two polynomial $q$-difference operators are given by
\begin{gather}\label{qGoursat2}
P_r\big(\sigma_q(x)\big)=\big(-\sigma_q(x)\big)^\mu \prod _{j=1}^r \big(1-a_j \sigma_q(x)\big)\,; \\
Q_s\big(\sigma_q(x)\big)=\left(-\frac{\sigma_q(x)}{q}\right)^m \prod _{k=0}^s \left(1-\frac{b_k}{q}\sigma_q(x)\right).
\end{gather}
Here $b_0=q$ and $m,\mu$ are constrained to satisfy 
\be
m+s+1 \leq r \quad\text{ and }\quad m+s+1+\mu=r\,. 
\ee
The solutions to this general equation are the basic $q$-hypergeometric series 
\be
f(x)={}_r\phi_s(a_1,...,a_r; b_1,..,b_s; q, x)\quad\text{for }\; 0\leq s\leq r\,, 
\ee
which have the following power series expansion for $|x|<1$:
\begin{equation}
{}_r\phi_s(a_1,...,a_r; b_1,..,b_s; q, x) =\sum _{n\geq 0}\frac{(a_1,...,a_r;q)_n}{(b_1,..,b_s ;q)_n (q;q)_n}\left[(-1)^n q^{\frac{n}{2}(n-1)}\right]^{1+s-r} x^n\,.
\end{equation}

\section{Special functions}\label{specialfunctions}

\begin{itemize}
\item $q$-Pochhammer symbol (finite case):
\begin{equation}
(x;q)_n =\prod _{i=0}^{n-1}(1-q^i x)  \,.
\end{equation}
\item $q$-Pochhammer symbol (infinite case):
\begin{align}
\label{qPochinfinite}
 (z;q)_\infty  &= \begin{cases}
 \prod _{n=0}^\infty (1-zq^n) & \text{for } |q|<1 \\
 \prod _{n=1}^\infty (1-zq^{-n})^{-1} & \text{for } |q|>1 \end{cases} \\
 &=\sum _{n=0}^\infty \frac{(-1)^n q^{\frac{1}{2}n(n-1)}}{(q;q)_n} z^n \,.
\end{align}
A useful identity is the inversion formula:
\be
(x;q)_\infty=\frac{1}{(q^{-1}x;q^{-1})_{\infty}} \,\cdot
\ee

\item $q$-Jacobi~theta function: 
\begin{align}
\Theta_q (x) &= (x;q)_\infty (qx^{-1};q)_\infty (q;q)_\infty \equiv (x,qx^{-1},q;q)_\infty\,,  \label{qthetafunction} 
\end{align}
with $x \in \mathbb{C} $  for $|q|<1$ and $x \in \mathbb{C} \setminus q^{\mathbb{Z}}$ for $|q|>1$.
Using Jacobi triple product formula, $q$-Jacobi theta function has the following series expansion:
\begin{equation}
\Theta _q (x)= \begin{cases} \sum _{n \in \mathbb{Z}}(-1)^n q^{\frac{n}{2}(n-1)}x^n & \text{for } |q|<1 \\
\left(\sum _{n \in \mathbb{Z}}(-1)^n q^{-\frac{n}{2}(n+1)}x^n\right)^{-1} & \text{for } |q|>1\,. \end{cases}
\label{qthetafunction1}
\end{equation}

\item The $\mathcal{J}(x,y;q)$ function is defined in terms of ${}_1 \phi _1$ $q$-hypergeometric function as follows:
\begin{align}
\mathcal{J}(x,y;q)&=(qy;q)_\infty {}_1 \phi _1(0;qy;q,qx)\,.
\end{align}
The series expansion for $|x|<1$ is given by
\begin{align}
\label{Jexpansion}
\mathcal{J}(x,y;q)=(qy;q)_\infty \sum _{n \geq 0} \frac{(-1)^n q^{\frac{n}{2}(n+1)}}{(qy,q;q)_n}x^n=(qy;q)_\infty \sum _{n \geq 0} \frac{x^n}{(q^{-1};q^{-1})_n(qy;q)_n}\,\cdot
\end{align}
Following are some useful identities that follow directly from the above series expansion:
\begin{align}
\text{For }|q| \neq 1: &\qquad \mathcal{J}(x,y;q)=\Theta _q (qy)\mathcal{J}(xy^{-1},y^{-1};q^{-1})  \label{A.7-1} \\
\text{For }|q|<1: &\qquad \mathcal{J}(x,y;q)= \mathcal{J}(y,x;q)  \label{Jqlessone} \\
\text{For }|q|>1: &\qquad \Theta_q (qx^{-1}y) \mathcal{J}(x,y;q)= \Theta _q (qy)\mathcal{J}(x^{-1},x^{-1}y;q)\,. \label{Jqmoreone}
\end{align}

\end{itemize}

\section{\texorpdfstring{Derivation of the connection formula for $\bm{|q| >1}$}{Derivation of the connection formula for |q|>1}}\label{proofofconn}

We start with the $q$-difference equation
\be
[a t\sigma _q^{-1}(t) +(1-(a+q)t) +qt\sigma _q(t)]h(t)=0
\label{master}
\ee
and substitute the following ansatz:
\be
h(t)=\frac{1}{\Theta _q(aqt)} u(t)\,,
\label{solutionh1app}
\ee
to get a $q$-difference equation for $u(t)$:
\be
\left[aq^2t^2 \sigma _q(t) +\left\lbrace (a+q)t-1\right\rbrace +\sigma_q^{-1}(t)\right] u(t)=0\,.
\label{scaledeqn}
\ee
For $|q|>1$, the $q$-Borel transform $\mathcal{B}^-_q$ of the above equation has a divergent $q$-Laplace transform. To tackle this divergence, we follow the general strategy in \cite{DRELO, OHY} and consider the order-$\frac{1}{2}$ $q$-Borel transform. Thus, we define
\be
q=p^2\,.
\ee
As we shall see, while the general methods of \cite{OHY} are used, there are some important differences in the details of this analysis for $|q|>1$. We first rewrite the $q$-difference equation \eqref{scaledeqn} as a $p$-difference equation:
\begin{align}
\left[ap^4t^2 \sigma_p^2(t) +\left(\big(a+p^2\big)t-1\right) +\sigma_p^{-2}(t)\right] u(t)=0\,.
\end{align}
Then using \eqref{Bplusminusids}, we arrive at the difference equation satisfied by the $p$-Borel transformed $u(t)$:
\be
\left[ (ap^7 \tau ^2 -1) \sigma _p^2(\tau) +(a+p^2)p^2 \tau \sigma_p(\tau) +1 \right]\widetilde{u}(\tau) =0\,,
\ee
where we have defined $\widetilde{u}(\tau) = \mathcal{B}_p^{-}[u(t)](\tau)$.

Now the main insight of \cite{OHY} is to transform this $p$-difference equation to the $q$-Goursat equation for the ${}_2 \phi_1$-function and use Watson's connection formula in order to obtain the connection formula for the ${}_1\phi_1$-function. We skip the intermediate steps for the former part and directly write down the solution for this $p$-difference equation: 
\be
\widetilde{u}(\tau)= (p^2 \alpha \tau ;p)_\infty~ {}_2 \phi _1 \left(\frac{p^2}{\alpha},\frac{\alpha}{p} ;-p; p ,p^2 \alpha \tau\right),
\ee
where we have defined $\alpha = \sqrt{ap}$. At this point, we would like to exploit the results of \cite{OHY} that are valid in the $|q|<1$ chamber. In order to do so, we define $p = Q^{-1}$ and write $\widetilde{u}(\tau)$ in terms of $Q$ such that we are in the $|Q|<1$ chamber. Using the inversion theorem for the Pochhammer symbol and the following identity for ${}_2\phi_1$: 
\be
{}_2 \phi _1 (a,b;c;p,z)={}_2 \phi _1 \left(\frac{1}{a},\frac{1}{b};\frac{1}{c};\frac{1}{p},\frac{abz}{cp}\right),
\ee
we get
\be
\widetilde{u}(\tau)=\frac{1}{(Q^{-1}\alpha \tau ;Q)_\infty}~{}_2 \phi _1 \big(\alpha Q^2 ,(Q \alpha)^{-1};-Q;Q,-Q^{-1}\alpha \tau\big)\,.
\ee

We now use Watson's formula (for $|Q|<1$) to find the connection formula:  
\be
\widetilde{u}(\tau) \longrightarrow \widetilde\gamma_1\, {}_2 \phi _1 \left(\alpha Q^2,-\alpha Q^2;\alpha ^2 Q^4;Q,\frac{Q^2}{\alpha \tau}\right) +\widetilde\gamma_2\, {}_2 \phi _1 \left(\frac{1}{Q\alpha},\frac{-1}{Q\alpha};\frac{1}{Q^2\alpha^2};Q,\frac{Q^2}{\alpha \tau}\right),
\label{u1tildeQ}
\ee
where the coefficients are given by
\begin{align}
\widetilde\gamma_1 &=\frac{(Q^{-1}\alpha ^{-1},-Q^{-1}\alpha ^{-1};Q)_\infty}{(-Q,Q^{-3}\alpha ^{-2};Q)_\infty}\frac{\Theta _Q(-Q \alpha ^2 \tau)}{\Theta _Q(-Q^{-1} \alpha \tau)(Q^{-1} \alpha \tau;Q)_\infty}\,; \\
\widetilde\gamma_2 &=\frac{(Q^{2}\alpha,-Q^{2}\alpha;Q)_\infty}{(-Q,Q^{3}\alpha ^{2};Q)_\infty}\frac{\Theta _Q(-Q^{-2}  \tau)}{\Theta _Q(-Q^{-1} \alpha \tau)(Q^{-1} \alpha \tau;Q)_\infty}\,\cdot
\end{align} 
In the $|Q|<1$ chamber, we have the following identity which relates the $q$-hypergeometric functions ${}_0 \phi _1$ and ${}_2 \phi _1$: 
\be
{}_2 \phi _1 \big(a,-a;a^2;Q,x\big) = \frac{1}{(x;Q)_\infty } {}_0 \phi _1 \big(-;a^2Q;Q^2,a^2Qx^2\big)\,. \label{identity}
\ee
Using this one rewrites \eqref{u1tildeQ} as 
\be
\widetilde{u}(\tau) \longrightarrow \frac{(Q^{-1}\alpha ^{-1},-Q^{-1}\alpha ^{-1};Q)_\infty}{(-Q,Q^{-3}\alpha ^{-2};Q)_\infty}{\mathcal W}_1(\tau)~ + \frac{(Q^{2}\alpha,-Q^{2}\alpha;Q)_\infty}{(-Q,Q^{3}\alpha ^{2};Q)_\infty}  ~{\mathcal W}_2(\tau)\,,
\label{borel2}
\ee
where we have separated out the $\tau$-dependent pieces into the ${\mathcal W}_i(\tau)$ that are given by
\begin{align}
{\mathcal W}_1(\tau) &=\frac{\Theta _Q(-Q \alpha ^2 \tau)}{\Theta _Q(-Q^{-1} \alpha \tau)\Theta _Q(Q^{-1} \alpha \tau)}{}_0 \phi _1 \big(-;\alpha ^2 Q^5;Q^2,Q^9\tau ^{-2}\big)  \label{coeff1}\,; \\
{\mathcal W}_2(\tau) &=\frac{\Theta _Q(-Q^{-2}  \tau)}{\Theta _Q(-Q^{-1} \alpha \tau)\Theta _Q(Q^{-1} \alpha \tau)} {}_0 \phi _1 \big(-;\alpha ^{-2} Q^{-1};Q^2,Q^3 \alpha ^{-4} \tau ^{-2}\big)\,.
\label{coeff2}
\end{align}
We now recall the following important lemma proved in \cite{OHY}:\footnote{We note that $\theta_q(x)|_{\text{there}}=\Theta _q(-x)|_{\text{here}}$.} Given 
\be
\phi (\tau)=\frac{\Theta _Q(-a\tau)}{\Theta _Q(-b_1\tau)\Theta _Q(-b_2\tau)}\sum _{m \geq 0}C_m\tau^{-2m}\,,
\ee
its $Q$-Laplace transform reads
\be
\mathcal{L}^+_{Q,\lambda} [\phi (\tau)](t)=\frac{\Theta _Q(-a\lambda)\Theta _{Q^{2}}(-aQ^2tb_1^{-1}b_2^{-1}\lambda^{-2})}{\Theta _Q(-b_1 \lambda)\Theta _Q(-b_2 \lambda)\Theta _Q(-Qt\lambda^{-1})} \sum _{m \geq 0} C_m Q^{-m(m-1)}\left(\frac{b_1b_2}{aQ^2 t}\right)^m.
\label{lemma1}
\ee
It is important here to note that the Laplace transform is being done w.r.t the variable $Q = p^{-1} = q^{-\frac{1}{2}}$ and the $+$ transform has been used. This is an important change from the $|q|<1$ case discussed in \cite{OHY} in which the Laplace transform is done w.r.t $q^{+\frac{1}{2}}$. The idea now is to apply the $Q$-Laplace transform operator sequentially to the two terms in the $Q$-Borel transformed solution in \eqref{borel2}. Let us begin with the first term, which can be expanded as:
\be
\mathcal{W}_1(\tau) =\frac{\Theta _Q(-Q \alpha ^2 \tau)}{\Theta _Q(-Q^{-1} \alpha \tau)\Theta _Q(Q^{-1} \alpha \tau)} \sum _{n \geq 0} \frac{Q^{2n^2 +7n}}{(\alpha ^2 Q^5,Q^2;Q^2)_n}\tau ^{-2n} \,.
\ee
The $Q$-Laplace transform can be computed using \eqref{lemma1}: 
\begin{align}
\mathcal{L}^+_{Q,\lambda} [\mathcal{W}_1 (\tau)](t) &=\frac{\Theta _Q(-Q\alpha ^2 \lambda)\Theta _{Q^2}(Q^5t\lambda^{-2})}{\Theta _Q(-Q^{-1}\alpha \lambda)\Theta _Q(Q^{-1}\alpha \lambda)\Theta _Q(-Qt\lambda^{-1})} 
\sum _{m \geq 0} \frac{(-1)^m(Q^2)^{\frac{m}{2}(m-1)}}{(\alpha ^2 Q^5,Q^2;Q^2)_m}\big(Q^4 \tau^{-1}\big)^m \nonumber\\
&= \frac{\Theta _Q(-Q\alpha ^2 \lambda)\Theta _{Q^2}(Q^5t\lambda^{-2})}{\Theta _Q(-Q^{-1}\alpha \lambda)\Theta _Q(Q^{-1}\alpha \lambda)\Theta _Q(-Qt\lambda^{-1})} ~{}_1 \phi _1 \big(0;\alpha ^2 Q^5;Q^2,Q^4\tau^{-1}\big)\,.
\end{align} 
This result is valid in the $|Q|<1$ chamber. We are eventually interested in writing the $Q$-Borel resummed solution in which the ${}_1\phi_1$-function has $q=p^2$ as the $q$-parameter. To that end, we make use of the following inversion formula:
\be
{}_1 \phi _1 (0;y;q,\tau)={}_1 \phi _1 \left(0;\frac{1}{y};\frac{1}{q},\frac{\tau}{qy}\right)
\label{conversionform}
\ee
so that the $Q$-Laplace transform can be written as:
\be
\mathcal{L}^+_{Q,\lambda} [\mathcal{W}_1 (\tau)](t) =\frac{\Theta _Q(-Q\alpha ^2 \lambda)\Theta _{Q^2}(Q^5t\lambda^{-2})}{\Theta _Q(-Q^{-1}\alpha \lambda)\Theta _Q(Q^{-1}\alpha \lambda)\Theta _Q(-Qt\lambda^{-1})} ~{}_1 \phi _1 \left(0;\frac{1}{Q^5\alpha ^2};\frac{1}{Q^2},\frac{1}{Q^3\alpha ^2 t}\right).
\ee 
Using $\Theta_{Q^2}(\tau)=[\Theta _{Q^{-2}}(Q^2\tau)]^{-1}$ and $p=Q^{-1}$, we obtain:
\be\label{LplusonW1}
\mathcal{L}^-_{p,\lambda} [\mathcal{W}_1 (\tau)](t) = \frac{\Theta _p(-p^2 \alpha \lambda)\Theta _p(p^2 \alpha \lambda)\Theta _p(-p \lambda t^{-1})}{\Theta _p(-\alpha ^2 \lambda)\Theta _{p^2}(\lambda ^2 p^5 t^{-1})}~{}_1 \phi _1 \left(0;\frac{p^{5}}{\alpha ^2};p^{2},\frac{p^{3}}{\alpha ^2 t}\right).
\ee
On the l.h.s we have changed notation by recalling that ${\mathcal L}^+_{p^{-1}, \lambda} ={\mathcal L}^-_{p, \lambda} $. This follows from the definitions in Section \ref{qborel}. We now substitute $\alpha^2 =ap$ and use $p^2=q$ to write the $q$-hypergeometric function in terms of $q$.
However, to simplify the $\Theta_p$ prefactors above, we need the following identities:
\be
\Theta _{q} (x)\Theta _{q} (px)=\Theta _p(x) \quad\text{ and }\quad \Theta _p(x)\Theta _p(-x)=\Theta _{q} (x^2)\label{theta}\,.
\ee
We also need to work on the Pochhammer symbols appearing as coefficient of $\mathcal{W}_1$ in \eqref{borel2}. To simplify those, we use the following identities: 
\be
(a,-a;q)_\infty =(a^2;q^2)_\infty \quad\text{ and }\quad (a;q)_\infty =(a;q^2)_\infty (aq;q^2)_\infty\,,
\label{pochhammer}
\ee
which lead to the following simplified coefficient:
\be
\frac{(-p,p^4 \alpha ^{-2};p)_\infty}{(p^2 \alpha ^{-1},-p^2 \alpha ^{-1};p)_\infty} =(-p;p)_\infty(p^4 a^{-1};p^2)_\infty =(-p;p)_\infty(q^2 a^{-1};q)_\infty \,.
\ee
Interestingly, the $q$-Pochhammer above can be combined with the ${}_1\phi_1$-function in \eqref{LplusonW1} to write it as a ${\mathcal J}$-function. Thus, combining all these factors, one can write the contribution to the analytic continuation that arises from the ${\mathcal W}_1$ term as:
\be
\frac{\Theta _{q}(apq^2 \lambda ^2)\Theta _{q} (-p\lambda t^{-1})\Theta _{q} (-q\lambda t^{-1})}{\Theta _{q}(-ap\lambda)\Theta _{q}(-ap^2 \lambda)\Theta _{q}(\lambda^2 p^5 t^{-1})} \mathcal{J} \left(a^{-1}t^{-1},q a^{-1};q\right).
\label{result1}
\ee
The term proportional to ${\mathcal W}_2$ can be analyzed along the same lines and we can finally write down the connection formula for $h(t)$ in the $|q|>1$ chamber (now taking into account the inverse $\Theta_q(aqt)$ factor in \eqref{solutionh1app}):
\begin{multline}
h(t) \longrightarrow \frac{\Theta _{q}(apq^2 \lambda ^2)\Theta _{q} (-p\lambda t^{-1})\Theta _{q} (-q\lambda t^{-1})}{\Theta _{q}(aq t)\Theta _{q}(-ap \lambda)\Theta _{q}(-ap^2 \lambda)\Theta _{q}(\lambda^2 p^5 t^{-1})} \mathcal{J} \left(a^{-1}t^{-1},q a^{-1};q\right) \\
+\frac{\Theta _{q}(a^{-1}p^{-3}\lambda ^{-2})\Theta _{q}(-p\lambda t^{-1})\Theta _{q}(-q\lambda t^{-1})}{\Theta _{q}(aq t)\Theta _{q}(-q^{-1}\lambda^{-1})\Theta _{q}(-p^{-1}\lambda^{-1})\Theta _{q}(apq\lambda^2 t^{-1})} \mathcal{J} \left(q^{-1}t^{-1},q^{-1} a;q\right).
\end{multline}
This is the result we used in subsection \ref{newconfrm} in equation \eqref{connformqmore1}.

\end{appendix}

\bibliographystyle{hephys}
\bibliography{q-exact_bib}

\end{document}